\documentclass[aps,prl,superscriptaddress,showpacs,nofootinbib,notitlepage,onecolumn]{revtex4-1}
\usepackage{graphicx,subfigure,amsmath,amsfonts,amssymb,multirow,mathrsfs,threeparttable}
\usepackage[colorlinks,linkcolor=blue,citecolor=blue,urlcolor=blue]{hyperref}

\def\nat{Nature\ }
\def\aap{Astron.\ Astrophys.\ }
\def\apj{Astrophys.\ J.\ }
\def\apjl{Astrophys.\ J.\ Lett.\ }

\def\mnras{Mon.\ Not.\ Roy.\ Astron.\ Soc.\ }

\def\prd{Phys.\ Rev.\ D\ }
\def\prl{Phys.\ Rev.\ Lett.\ }

\begin{document}
\newcommand{\PMO}{Key Laboratory of Dark Matter and Space Astronomy, Purple Mountain Observatory, Chinese Academy of Sciences, Nanjing 210033, People's Republic of China.}
\newcommand{\USTC}{School of Astronomy and Space Science, University of Science and Technology of China, Hefei, Anhui 230026, People's Republic of China.}
\newcommand{\AHNU}{Department of Physics, Anhui Normal University, Wuhu, Anhui 241000, People's Republic of China.}
\newcommand{\GXU}{Guangxi Key Laboratory for Relativistic Astrophysics, Nanning 530004, People's Republic of China.}

\title{Mass and effective spin distributions of binary black hole events: Evidence for a population of hierarchical black hole mergers in active galactic nuclei disks}

\author{Yuan-Zhu Wang}
\affiliation{\PMO}
\author{Yi-Zhong Fan}
\email{yzfan@pmo.ac.cn}
\author{Shao-Peng Tang}
\affiliation{\PMO}
\affiliation{\USTC}
\author{Ying Qin}
\email{yingqin2013@hotmail.com}
\affiliation{\AHNU}
\affiliation{\GXU}

\author{Da-Ming Wei}
\affiliation{\PMO}
\affiliation{\USTC}
\date{\today}

\begin{abstract}
The origins of the coalescing binary black holes (BBHs) detected by the advanced LIGO/Virgo are still in debate 
and clues may present in the mass and effective spin ($\chi_{\rm eff}$) distributions of these merger events. Here we analyze the GWTC-1 and GWTC-2.1 BBH events, and find evidence for a primary mass dependent $\chi_{\rm eff}$ distribution. Below (above) the mass of $\sim 50M_\odot$, a nearly zero (relatively high) effective spin is typical. 
In the joint analysis of the primary mass and $\chi_{\rm eff}$ distributions, we identify a sub-population of BBHs characterized by both a bump-like primary mass distribution extending to $\sim 100M_\odot$ and a relatively high $\chi_{\rm eff}\sim 0.3$, consistent with the expectation for the $\geq 2$g merger events in the disks of active galactic nuclei (AGNs). 
Our results suggest that the merger rate in AGN disks is around $1.7~{\rm Gpc}^{-3}{\rm yr}^{-1}$.
\end{abstract}
\maketitle

{\it Introduction.---} With the help of the Advanced LIGO and Virgo detector network, the number of observed binary black hole (BBH) mergers is rapidly growing. To date, more than 50 BBH merger candidates have been released with the current catalogs of compact binary coalescences (GWTC-1 and GWTC-2.1) \cite{abbottO2,2021arXiv210801045T}.
The origins of these compact objects are still unclear. Various formation channels have been proposed, including for instance the isolated binary evolution, dynamical capture, and the AGN enhancement (see Refs.~\cite{mapelli2021review,gerosa2021review} for recent reviews). 

Formation channels may leave their ``finger prints" in the final population of BHs. Of all various channels, the leading scenario for the progenitor of BBH is a close binary system composed of a BH and a helium star, which can be the outcome of the classical isolated binary evolution \cite{2016Natur.534..512B} through the common envelope \cite{2013A&ARv..21...59I}. From the point of view of stellar evolution theory, the BH masses are expected to fall outside a gap starting at $\sim 40-65 M_\odot$ and ending at $\sim 125M_\odot$ due to (pulsational) pair-instability supernovae ((P)PISNe) \cite{2016A&A...594A..97B,2017ApJ...836..244W,2017MNRAS.470.4739S,2019ApJ...882..121S,2020ApJ...888...76M}. The BH spins are generally small \cite{2018A&A...616A..28Q,2019MNRAS.485.3661F,2020A&A...636A.104B} under the assumption of efficient angular momentum transport within stars. On the other hand, in the chain of hierarchical mergers, BHs inside the PISN/PPISN mass gap can appear in the second or higher-generation mergers, and large spin magnitudes for the primary BHs are expected \cite{miller2002,giersz2015,fishbach2017,gerosa2017,rodriguez2019,arcasedda2021b,mapelli2021,gerosa2021review}. More specifically, for the hierarchical mergers happened in the AGN disk, BHs may align their orbits with the disk, leading to a broad effective spin ($\chi_{\rm eff}$) distribution \cite{2019PhRvL.123r1101Y,2021MNRAS.507.3362T}.

Population studies were carried out with analytical models guided by theoretical concerns \cite{2020ApJ...900..177K,2021ApJ...913L...7A,2021ApJ...913...42W}, simulations \cite{2021ApJ...916L..16B,2020A&A...635A..97B} or non-parametric approaches \cite{2021ApJ...917...33L,2021arXiv210905960R} to recover the astrophysical distributions of BBHs. The features/substructures in the primary mass function \cite{2021ApJ...913L...7A,2021ApJ...913...42W,2021CQGra..38o5007T,2021ApJ...917...33L} and potential correlations \cite{2021arXiv210600521C,2021ApJ...907L..24S,2021arXiv210708811T} may provide additional clues for their formation histories. The observed events are likely originated from multiple channels \cite{2020ApJ...900..177K,2021ApJ...913...42W,bouffanais2021,zevin2021,wong2021,roulet2021}.To unveil the physical origin, in this work we search for additional clues for sub-populations with different origins in the spin distribution. We focus on the well-measured effective spin parameter ($\chi_{\rm eff}$) and aim to construct simple analytical models with astrophysical motivations to describe the mass and spin distributions self-consistently. We find out that the properties of inferred nonstellar-origin sub-population are consistent with that predicted in Ref.~\cite{2019PhRvL.123r1101Y}, in favor of the presence of a population of hierarchical black hole mergers in AGN disks. 

\begin{figure}[htbp]
	\centering
	\includegraphics[angle=0,scale=0.575]{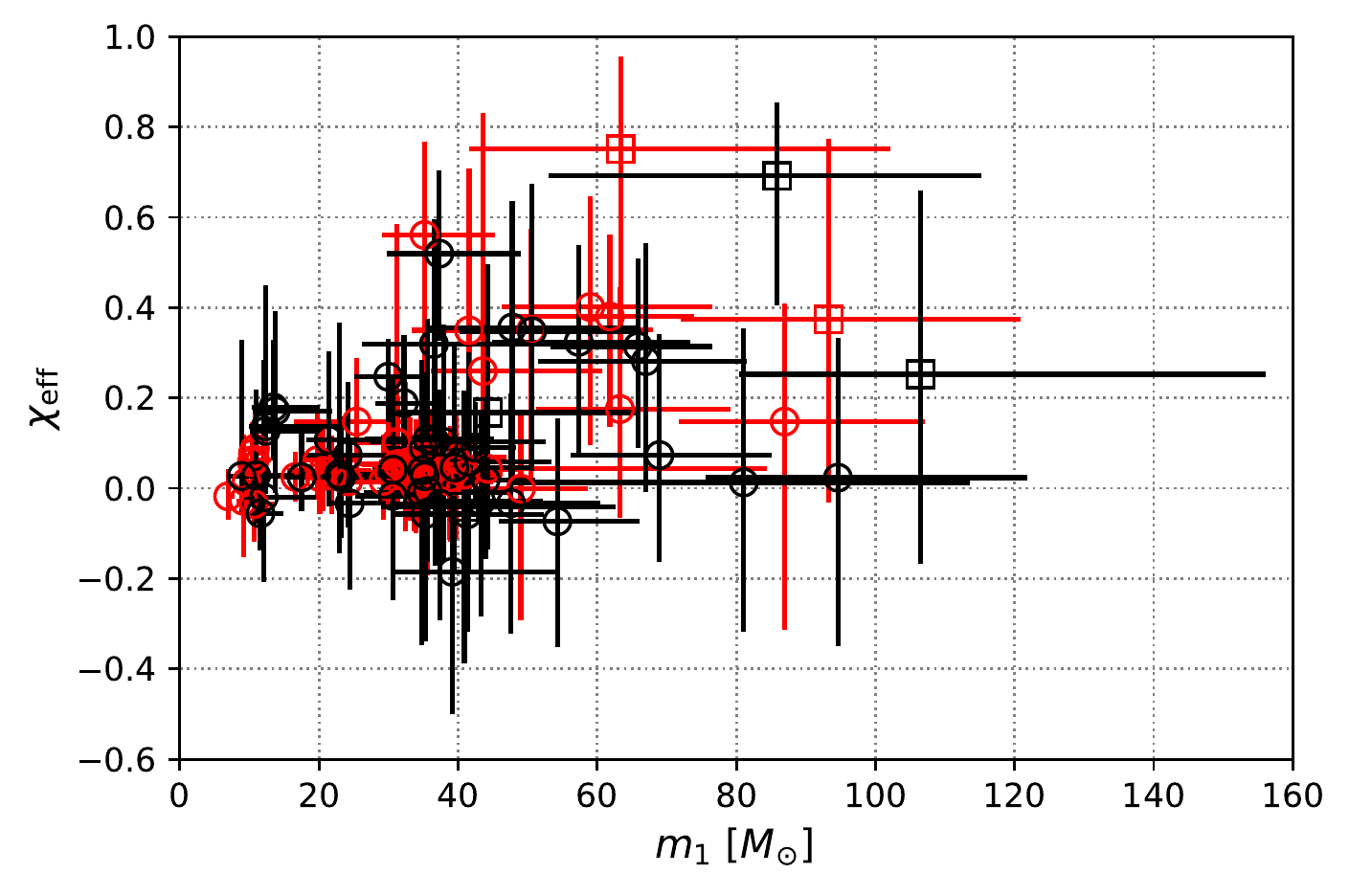}
	\includegraphics[angle=0,scale=0.56]{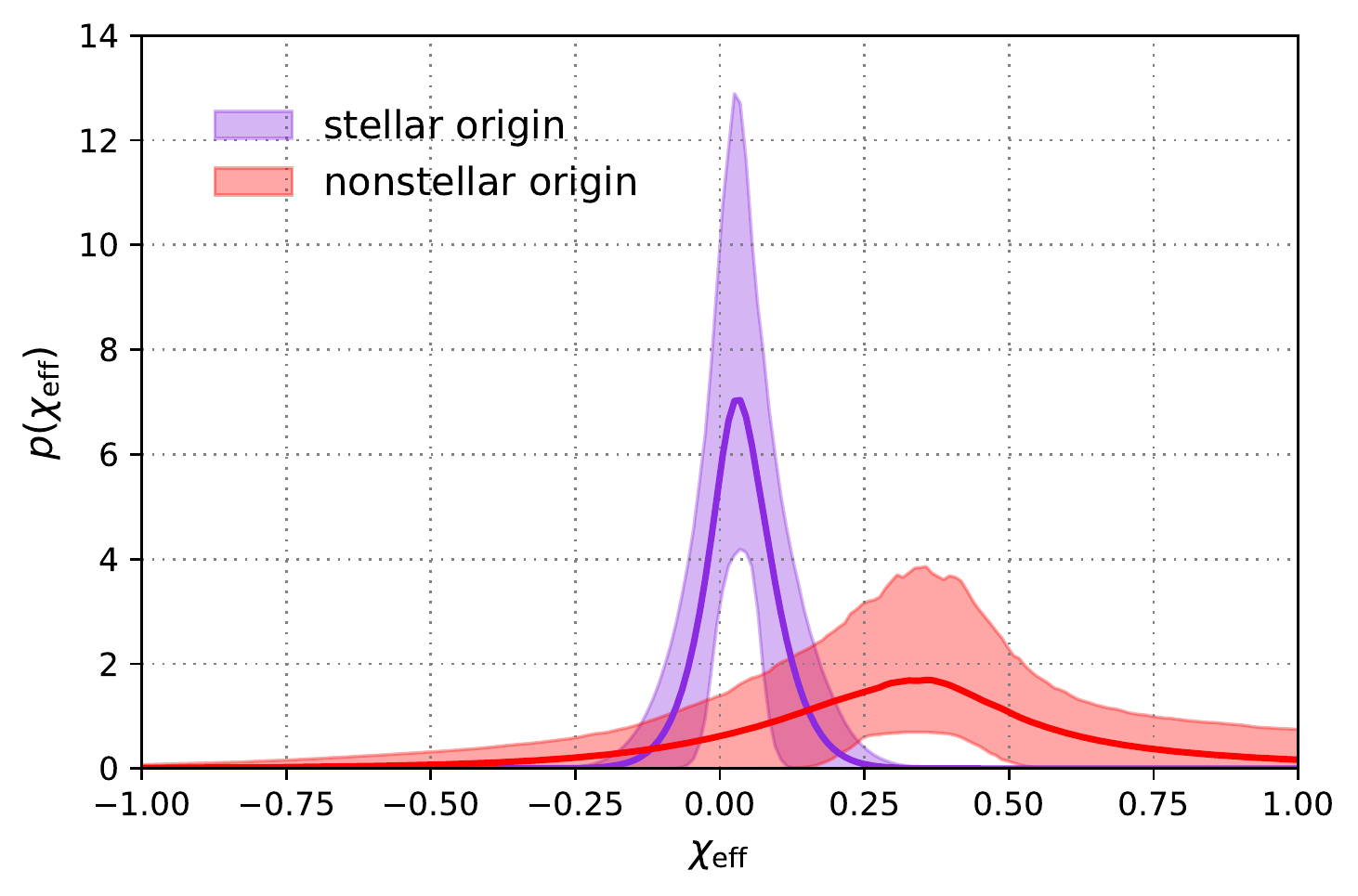}
	\caption{Left panel: Observed effective spin and primary mass of each event. The black data points are the results inferred with the uninformative default prior, while the red data points are the re-weighted results informed by our population model. The circle (squared) markers represent the events with FAR$\leq 1/{\rm yr}$ ($>1/{\rm yr}$). The error bars show the $90\%$ credible intervals. Right panel: Reconstruction of the two sub-populations in our spin model. The shaded areas are the $90\%$ credible bounds, and the solid curves are the posterior population distribution averaging over model uncertainty.}
	\hfill
\label{fig:spin}
\end{figure}

{\it Population Models.---} Our spin model is inspired by the potential transition on the $\chi_{\rm eff} - m_1$ plane of the observation data \cite{abbottO2,2021arXiv210801045T}. As shown in the left panel of Fig.~\ref{fig:spin}, the data are clustering at $\chi_{\rm eff} \sim 0$ for $m_1 \leq 40-60 M_\odot$, above which a typical $\chi_{\rm eff}\sim 0.2-0.4$ is found. Since this transition mass is consistent with the theoretical prediction on the lower edge of PPISN gap, we propose a $\chi_{\rm eff}$ distribution model consisting of two sub-populations: one has the stellar origin (either field binaries or dynamical capture of isolated stars), and the other is formed through for instance hierarchical merger.  The classification of the sub-populations, assumed to have Gaussian distributions, for a specific event is mainly based on the primary mass. 
The division mass $m_{\rm d}$ is defined as a free parameter. For $m_1 \leq m_{\rm d}$, the probability density function of $\chi_{\rm eff}$ is a mixture of the two sub-populations, and the fraction of stellar origin events is represented by $f_{\rm mix}$; otherwise it simply follows the second Gaussian distribution. We then have
\begin{equation}\label{eq:spin_pwg}
\begin{aligned}
    &\pi_{\rm s}(\chi_{\rm eff} \mid m_1, \mu_{\rm 1,eff},\sigma_{\rm 1,eff},\mu_{\rm 2,eff},\sigma_{\rm 2,eff}, m_{\rm d}, f_{\rm mix})\\= \, & f_{\rm mix}\mathcal{H}(m_{\rm d}-m_1) \mathcal{N}(\chi_{\rm eff} \mid \mu_{\rm 1,eff},\sigma_{\rm 1,eff}) +  [1-f_{\rm mix}\mathcal{H}(m_{\rm d}-m_1)] \mathcal{N}(\chi_{\rm eff} \mid \mu_{\rm 2,eff},\sigma_{\rm 2,eff}),
\end{aligned}
\end{equation}
where $\mathcal{H}$ is the Heaviside step function, $(\mu_{\rm 1,eff},\sigma_{\rm 1,eff})$ and $(\mu_{\rm 2,eff},\sigma_{\rm 2,eff})$ are the means and standard deviations of the first and second Gaussian components, respectively. We truncate and normalize the two Gaussian components to ensure $-1 \leq \chi_{\rm eff} \leq 1$. Specifically, under the assumptions of the efficient angular momentum transport in stellar evolution \cite{2002A&A...381..923S,2019MNRAS.485.3661F} and PPISN explosion \cite{2016A&A...594A..97B}, one would expect $\mu_{\rm 1,eff}\sim 0$  \cite{2018A&A...616A..28Q,2019MNRAS.485.3661F,2020A&A...636A.104B} and that $m_{\rm d}$ is consistent with the lower edge of PPISN gap. 

We adopt a primary mass distribution function from Ref.~\cite{2021ApJ...913...42W} (i.e., their model IV) which can better describe the LIGO/Virgo data than the fiducial POWER LAW+PEAK model suggested in Ref.~\cite{abbottO3popandrate}. This function also consists of sub-populations with stellar origin and nonstellar origin as our spin model. The stellar-origin component has a profile (modulated Power-Law + Gaussian) similar to the astrophysically motivated parametric model proposed by Ref.~\cite{2018ApJ...856..173T}, while the nonstellar-origin component is represented by a wide modulated Gaussian profile. The mass function reads
\begin{equation}\label{eq:plgg}
\begin{split}
\pi(m_1 \mid f_{\rm stellar}, f_{\rm PL}, \alpha, m_{\rm min}, \delta_m, m_{\rm max}, \mu_{\rm 1,m}, \sigma_{\rm 1,m}, \mu_{\rm 2,m}, \sigma_{\rm 2,m}) = f_{\rm stellar} f_{\rm PL} \mathcal{P}'(m_1 \mid \alpha,m_{\rm min}, \delta_m, m_{\rm max})\\ + f_{\rm stellar} (1-f_{\rm PL}) \mathcal{G}'(m_1 \mid m_{\rm min}, \delta_m, \mu_{\rm 1,m}, \sigma_{\rm 1,m}) + (1-f_{\rm stellar}) \mathcal{G}'(m_1 \mid m_{\rm min}, \delta_m,\mu_{\rm 2,m},\sigma_{\rm 2,m}),
\end{split}
\end{equation}
where $\mathcal{P}'$ and $\mathcal{G}'$ are the modulated power-law and modulated Gaussian distributions described in Ref.~\cite{2021ApJ...913...42W}, $f_{\rm stellar}$ is the fraction of events having stellar progenitors (among them, a fraction of $f_{\rm PL}$ is contributed from the modulated power-law component), $m_{\rm max}$ is the cut-off mass caused by PPISN, which should be identical to $m_{\rm d}$ in the spin model. 
 
{\it Methods.---} We perform hierarchical Bayesian inference to constrain the parameters. Our data set consists of BBH mergers reported with FAR$\leq 1/{\rm yr}$ so far \cite{2021arXiv210801045T,abbottO2}. In total our sample consists of 47 BBH events. 
Their posterior information is obtained from the literature~ \cite{abbottO2,abbottO3a,2021arXiv210801045T}. We adopt the ``Overall$_{-}$posterior'' samples for the events in GWTC-1, ``PublicationSamples'' for the events in GWTC-2, and ``IMRPhenomXPHM$_{-}$comoving'' samples for the new events in GWTC-2.1. The default parameter estimation priors of each individual event were also considered in the hierarchical inference to cancel out their influence on the posterior samples.

We first carry out the spin distribution analysis (i.e., SPIN ANALYSIS hereafter) following the procedures described in Ref.~\cite{2021ApJ...913L...7A} (the Appendix~D therein). Their fixed TRUNCATED mass model is adopted and a constant merger rate is assumed \footnote{The cases of more complex mass models or a redshift evolving merger rate ($\propto (1+z)^{2.7}$) have also been tested, and our results are not significantly affected.}. The conditional prior for a set of posterior samples, $(\chi_{\rm eff},m_1,m_2)$, is then given by
\begin{equation}\label{eq:llh_spin}
\begin{aligned}
    &\pi(\chi_{\rm eff}, m_1,m_2 \mid \mu_{\rm 1,eff},\sigma_{\rm 1,eff},\mu_{\rm 2,eff},\sigma_{\rm 2,eff}, m_{\rm d}, f_{\rm mix}, \mathbf{\Lambda}_{\rm m}) \\ = \, &\pi_{\rm s}(\chi_{\rm eff} \mid m_1, \mu_{\rm 1,eff},\sigma_{\rm 1,eff},\mu_{\rm 2,eff},\sigma_{\rm 2,eff}, m_{\rm d}, f_{\rm mix}) \; \pi_{\rm m}(m_1,m_2 \mid \mathbf{\Lambda}_{\rm m}),
\end{aligned}
\end{equation}
where $\pi_{\rm m}(m_1,m_2 \mid \mathbf{\Lambda}_{\rm m})$ is the TRUNCATED mass model with fixed parameters $\mathbf{\Lambda}_{\rm m}$. We incorporate Eq.~(\ref{eq:llh_spin}) into the likelihood for the hyper-parameters in the hierarchical inference (see Ref.~\cite{2019PASA...36...10T} for the expression of the likelihood), and use the python package {\sc Bilby} and {\sc MultiNest} sampler to obtain the Bayesian evidence and posteriors of the hyper-parameters.

Three priors for the hyper-parameters in Eq.~(\ref{eq:spin_pwg}) are used in our analyses, namely the Prior S1 (the fiducial prior), S2, and S3, which are summarized in Tab.~\ref{tb:metainfo}. The latter two priors have smaller prior volumes, and their adoptions will simplify the spin model. The comparison between the inferred Bayesian evidences for Prior S1 and S2 can reveal whether the spin distribution has a dependence on primary mass or not; the comparison between Prior S1/S2 and S3 can examine the necessity of describing the spin distribution with two components. In this SPIN ANALYSIS, we do not consider the spin-induced selection effect, since its impact on the inferred $\chi_{\rm eff}$ distribution is likely to be very small \cite{2021ApJ...913L...7A, 2021arXiv210902424G}.

\begin{table}
    \begin{threeparttable}
	\centering \caption{Constraints on the parameters of the spin model under different priors and mass models.}
	\label{tb:metainfo}
	\begin{tabular}{|c|c|c|c|c|c|c|c|c|} 
		\hline \multirow{2}{*}{\bf{Parameter Name}} & \multicolumn{2}{c|}{S1} & \multicolumn{2}{c|}{S2} & \multicolumn{2}{c|}{S3} & \multicolumn{2}{c|}{Joint} \\ \cline{2-9}
		                                                                      & Priors    & Posteriors   & Priors    & Posteriors     & Priors    & Posteriors   & Priors    & Posteriors \\
		\hline\hline
		$\mu_{\rm 1,eff}$     & [-0.1,0.1]  & $0.04^{+0.04}_{-0.03}$ & [-0.1,0.1] & $0.06^{+0.03}_{-0.04}$  & 1 & --        & [-0.1,0.1] & $0.04^{+0.03}_{-0.03}$    \\
		$\sigma_{\rm1,eff}$  & [0.01,0.1] & $0.06^{+0.03}_{-0.03}$ & [0.01,0.1] & $0.07^{+0.03}_{-0.03}$  & 1 & --        & [0.01,0.1] & $0.06^{+0.03}_{-0.03}$    \\
		$\mu_{\rm 2,eff}$     & [-1,1] & $0.31^{+0.34}_{-0.20}$ & [-1,1] & $0.46^{+0.44}_{-0.56}$  & [-1,1]  & $0.09^{+0.04}_{-0.04}$ & [-1,1] & $0.32^{+0.35}_{-0.20}$ \\
		$\sigma_{\rm 2,eff}$  & [0.01,1] & $0.24^{+0.50}_{-0.15}$ & [0.01,1] & $0.37^{+0.54}_{-0.31}$  & [0.01,1] & $0.12^{+0.05}_{-0.04}$ & [0.01,1] & $0.24^{+0.51}_{-0.16}$ \\
		$m_{\rm d} ~ [M_\odot]$          & [30,65] & $48^{+11}_{-9}$        & 300 & --                      & 300 & --       & [30,65] & $46^{+13}_{-8}$         \\
		$f_{\rm mix}$                    & [0.8,1] & $0.93^{+0.06}_{-0.10}$ & [0.8,1] & $0.90^{+0.07}_{-0.08}$ & 0  & --       & [0.8,1] & $0.92^{+0.07}_{-0.09}$    \\
		\hline
		$\ln{\mathcal{B}}$ (FAR$\leq 1/{\rm yr}$)  & \multicolumn{2}{c|}{0} & \multicolumn{2}{c|}{-1.68} & \multicolumn{2}{c|}{-7.43} & \multicolumn{2}{c|}{--} \\ \cline{1-9}
		$\ln{\mathcal{B}}$ (all events)  & \multicolumn{2}{c|}{0} & \multicolumn{2}{c|}{-2.47} & \multicolumn{2}{c|}{-10.19} & \multicolumn{2}{c|}{--} \\		\hline\end{tabular}
\begin{tablenotes}
\item The logarithmic Bayes factors of different priors compared with Prior S1 are shown in the last two rows in the table. The constraints on the hyper-parameters for the mass model are consistent with those from Ref.~\cite{2021ApJ...913...42W}, except for ``$m_{\rm max}$", which is shown as ``$m_{\rm d}$" in the table. $\mu_{\rm 1,eff}$, $\sigma_{\rm 1,eff}$, and $m_{\rm d}$ are fixed to arbitrary values in Prior S3, since they are irrelevant to the inference when $f_{\rm mix}=0$. 
\end{tablenotes}
\end{threeparttable}
\end{table}

We also perform a JOINT ANALYSIS, which simultaneously infers the parameters for the mass model, the spin model, as well as the merger rate $R_{\rm BBH}$ (assumed to be constant in time). We replace the primary mass function in the TRUNCATED mass model in Eq.~(\ref{eq:llh_spin}) with Eq.~(\ref{eq:plgg}), and set the $m_{\rm max}$ in the mass model to be equal to $m_{\rm b}$ in the spin model. Besides, we take the mass-induced selection effect into account following Ref.~\cite{2021ApJ...913L...7A}, where the information from Ref.~\cite{2021arXiv210801045T} is adopted to better approximate the sensitivity of the searching pipeline used in GWTC-2.1. 

{\it Results.---} The results of the SPIN ANALYSIS are presented in Tab.~\ref{tb:metainfo}. Using Prior S3, we recover a single Gaussian distribution with $\mu_{\rm 2,eff} = 0.09^{+0.04}_{-0.04}$ and $\sigma_{\rm 2,eff} = 0.12^{+0.05}_{-0.04}$, which is consistent with that found before \cite{2021ApJ...913L...7A}. However, the Bayes evidence of Prior S3 is at least $\sim 2.5$ magnitudes lower than the others. Therefore, a single Gaussian distribution of $\chi_{\rm eff}$ is strongly disfavored. Similar evidence for the existence of two sub-populations, without investigating the mass dependence, was found in Refs.~\cite{roulet2021, 2021arXiv210902424G}. We obtain consistent constraints on the common parameters using Prior S1 and S2, despite that the division mass, $m_{\rm d}$, can only be inferred by adopting Prior S1. The logarithmic Bayes factor ($\ln{\mathcal{B}}$) of Prior S2 compared with Prior S1 is -1.68, indicating a moderate support for the presence of the dependence upon $m_{\rm d}$. To check the robustness of this result, we also include the events with FAR$ > 1/{\rm yr}$ in the data set and repeat the analysis. The resulting $\ln{\mathcal{B}}$ is -2.47, which indicates an even stronger preference. 

\begin{figure}[htbp]
	\centering
	\includegraphics[angle=0,scale=0.6]{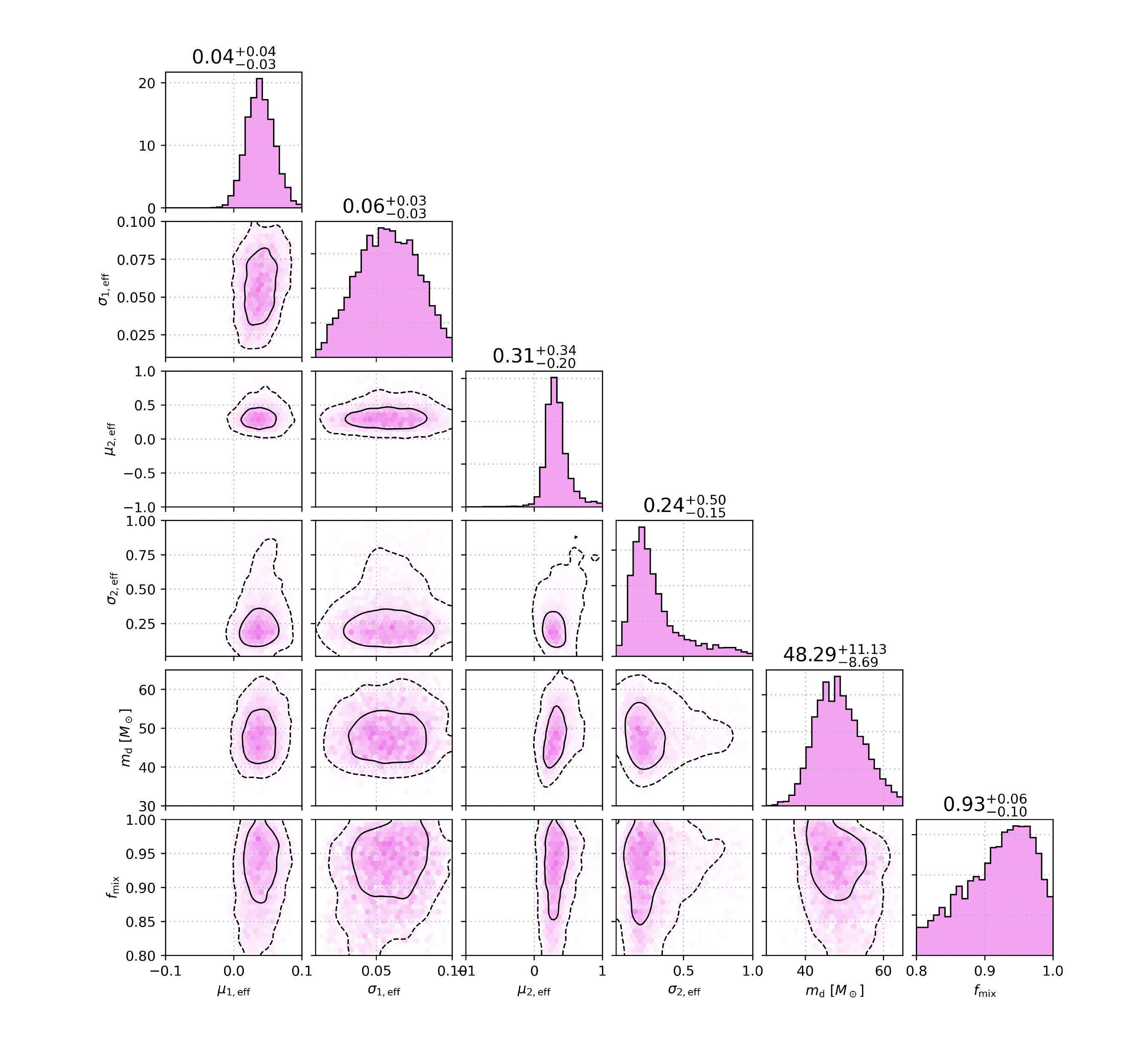}
	\caption{Posterior distributions for the parameters in the SPIN ANALYSIS inferred using Prior S1. The 50\% and 90\% credible regions are shown by the solid and dashed edges.}
	\hfill
\label{fig:PS1}
\end{figure}

The reconstruction of the two sub-populations in our spin model is shown in the right panel of Fig.~\ref{fig:spin}, and the posteriors for the parameters describing the corresponding model are presented in Fig.~\ref{fig:PS1}. The first Gaussian component (with stellar origin) centers at $\chi_{\rm eff}\approx 0$, while the second Gaussian component (with nonstellar origin) peaks at $\chi_{\rm eff}\approx 0.3$. The $m_{\rm d}$ is constrained to be $48^{+11}_{-9} M_\odot$, which is robust against the choice of a wider prior range (like $2-100 M_\odot$).

\begin{figure}[htbp]
	\centering
	\includegraphics[angle=0,scale=0.59]{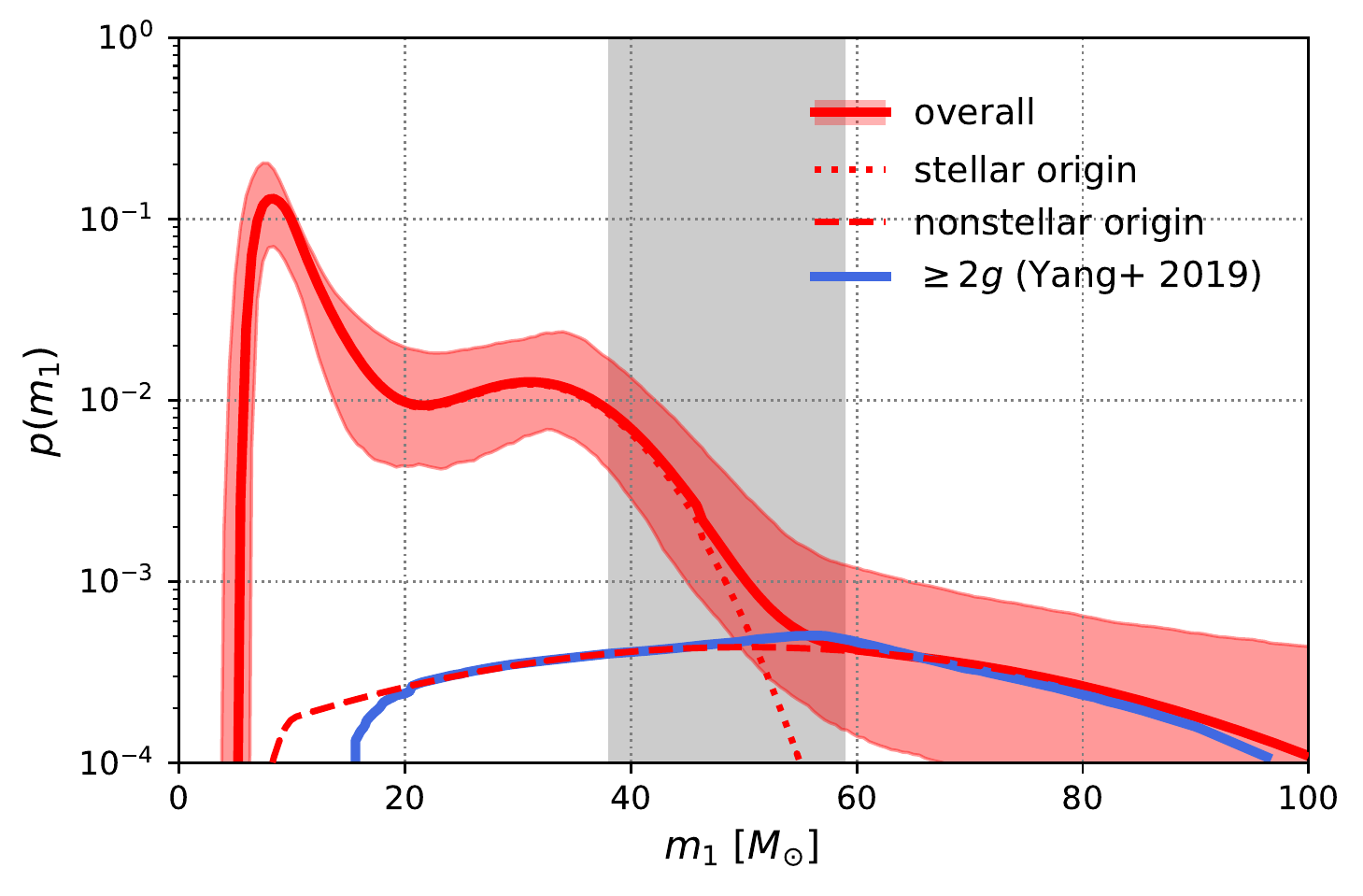}
	\includegraphics[angle=0,scale=0.59]{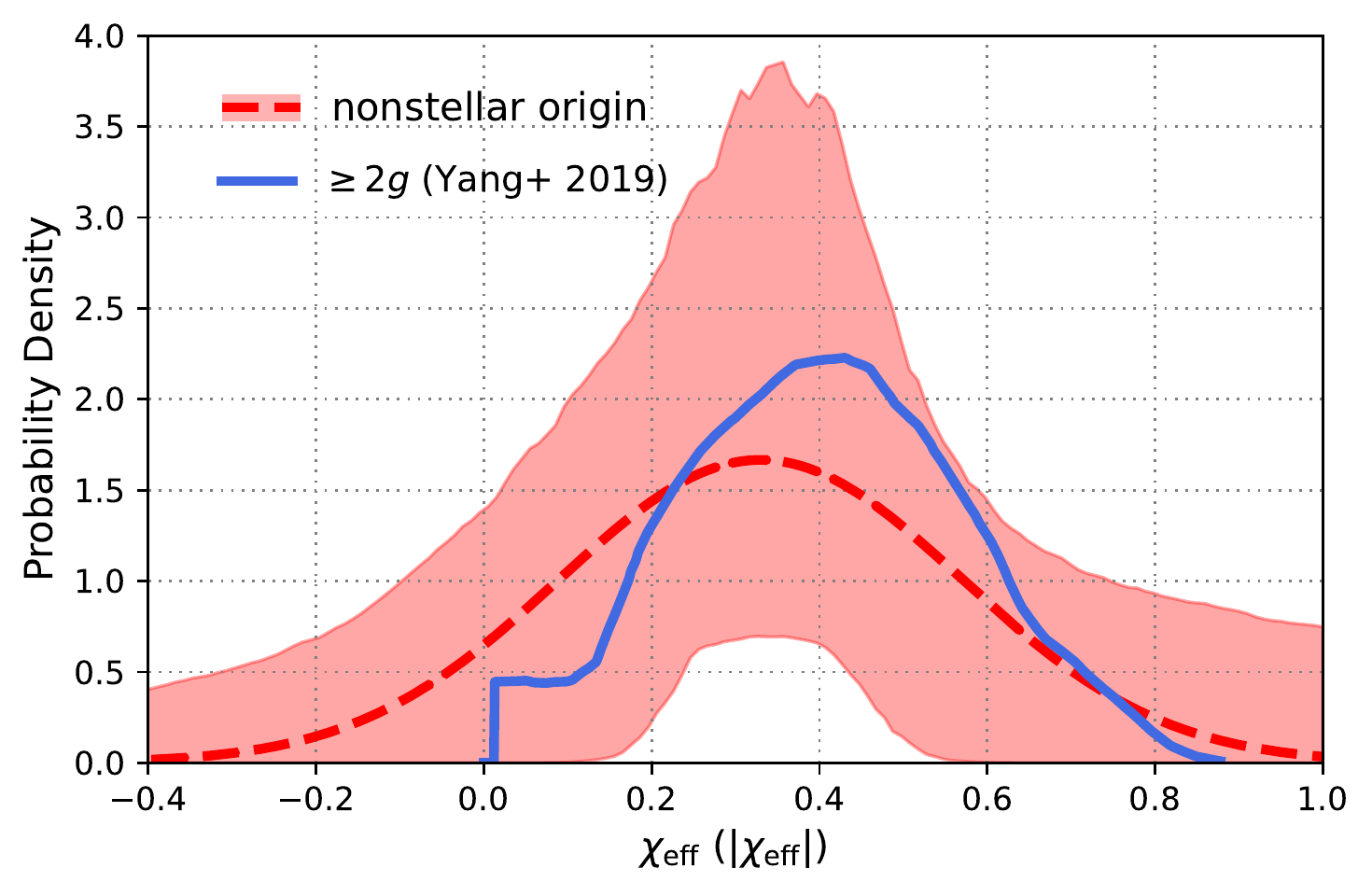}
	\caption{Reconstructions of primary mass distribution (left panel) and effective spin distribution for the sub-population with nonstellar origin (right panel) inferred from our joint analysis (red), comparing with the theoretical distribution for the hierarchical mergers in AGN disk (blue) \cite{2019PhRvL.123r1101Y}. The red shaded areas represent the $90 \%$ credible bounds. In the left panel, the red solid line shows a representative overall $m_1$ distribution, and its sub-components are marked with dotted and dashed lines; the blue solid line is the theoretical distribution produced by second and higher-generation mergers \cite{2019PhRvL.123r1101Y}. The grey shaded interval shows the constraint on $m_{\rm max}$ at $90 \%$ credibility, which overlaps the quickly-declining part of the Gaussian component. Therefore we conclude that there is a sharp drop in the mass function at $\sim 50M_\odot$. In the right panel, the red dashed line is a representative $\chi_{\rm eff}$ distribution for the sub-population with nonstellar origin.}
	\hfill
\label{fig:compare}
\end{figure}

For the JOINT ANALYSIS, the recovered mass and spin distributions are presented in Fig.~\ref{fig:compare}. The overall primary mass function extends to $\sim 100M_\odot$. One attractive feature is the rapid drop at $\sim 50 M_\odot$. The rapid drop is then followed by an emerging component, which is likely too massive to be of the stellar origin. In comparison to Ref.~\cite{2021ApJ...913...42W}, the hyper-parameters for the mass model are rather similar except that the constraint on $m_{\rm max}$ are improved. Such improvement is mainly due to the inclusion of the spin data. 
The JOINT ANALYSIS has enabled us to recover both mass and spin distributions, and these distributions can in turn be utilized as new population-informed priors in single event's parameter inference \cite{2020ApJ...891L..31F,2020ApJ...895..128M}. Following Ref.~\cite{2020ApJ...895..128M}, we re-weight the posteriors of each event, and the new posteriors informed by our population models are presented in the left panel of Fig.~\ref{fig:spin}. After the re-weighting, the difference of the $\chi_{\rm eff}$ distribution between low and high mass primary BHs becomes more conspicuous. 

{\it A population of BBH mergers from AGN disk?---} The JOINT ANALYSIS has revealed that the nonstellar-origin sub-population, characterized by a high effective spin, has an extended ``bump" like $m_1$ distribution. The hierarchical mergers are one of the most promising channels to produce similar populations \cite{2019PhRvL.123r1101Y, gerosa2021review}. The expected mass and spin distributions from hierarchical mergers may differ greatly for different considerations on the environments and detailed dynamical interactions \cite{2020ApJ...898...25T,2021MNRAS.507.3362T}. In the AGN scenario, BBHs can align their orbit with the disk, leading to large $|\chi_{\rm eff}|$ values for second or higher-generation mergers \cite{2019PhRvL.123r1101Y}. In addition, the preference of forming high mass primary BHs will significantly increase as the generation of mergers grows. 

We compare the reconstructed primary mass distribution with the theoretical distribution for the hierarchical mergers in AGN disk in the left panel of Fig.~\ref{fig:compare}. We show a representative primary mass distribution by fixing the hyper-parameters to their median values found in the modeling. The theoretical $m_1$ distribution predicted by the $\geq 2$g (``g" is short for ``generation") mergers is obtained by combining the distributions of different generations reported in Ref.~\cite{2019PhRvL.123r1101Y}. Then, by multiplying the normalized theoretical distribution with a factor of $0.03$, we find that the resulting curve is very similar to our nonstellar-origin component (see the red dashed line in left panel of Fig.~\ref{fig:compare}).
Giving that the fraction of $\geq 2$g mergers is $\approx 53\%$ in Ref.~\cite{2019PhRvL.123r1101Y} and the inferred $R_{\rm BBH}$ is $29.8^{+16.0}_{-9.8}~{\rm Gpc}^{-3}{\rm yr}^{-1}$, the merger rate for the AGN disk events can be roughly estimated as $\approx 1.7^{+0.9}_{-0.6}~{\rm Gpc}^{-3}{\rm yr}^{-1}$. The 1g AGN disk mergers contributes only a very limited portion ($\sim$ three percents) of the stellar origin component.

Similar comparisons are made for the $\chi_{\rm eff}$ distribution in the right panel of Fig.~\ref{fig:compare}. The blue curves in the figure are the $|\chi_{\rm eff}|$ distributions predicted by Ref.~\cite{2019PhRvL.123r1101Y}. Though both aligned and anti-aligned spins are possible for $\geq 2$g mergers, the aligned events can be dominant if the accretion onto the black holes is important \cite{2019ApJ...884L..12Y}. The representative distribution for the effective spin is extended to the negative $\chi_{\rm eff}$ region, however, as mentioned in Ref.~\cite{roulet2021}, the choice of empirical model may affect the interpretation on the existence of negative $\chi_{\rm eff}$.

Note that Ref.~\cite{2019PhRvL.123r1101Y} has used a power-law BH initial mass function (with a slope of $-2.35$ and the maximum mass of $50 M_\odot$) to approximate the 1g mergers in the AGN disks, which we think reasonable for the following reasons: i) the overall mass distribution of the stellar-origin component found in our modeling roughly matches the initial mass function adopted in Ref.~\cite{2019PhRvL.123r1101Y}; ii) their maximum mass for the power-law component is consistent with that of the stellar-origin objects, as found in this work. 

{\it Conclusion.---} We have analyzed the current LIGO/Virgo BBH events to investigate the physical origins of these merging black holes. Our conclusions are listed as follows: 
\begin{itemize}
\item The effective spin distribution is found to be dependent upon the primary masses of the BBH events. Below (above) the division mass of $m_{\rm d}\sim 50M_\odot$, a negligible (relatively high) effective spin is typical. 
In view of also the sharp drop of the primary mass distribution at $\sim 50M_\odot$, we suggest that the classical pulsational pair instability process does govern some supernova explosions. 

\item In the jointed analysis of the mass and effective spin distributions, we identify a sub-population of BBHs characterized by a bump-like primary mass distribution extending to $\sim 100M_\odot$ and a relatively high $\chi_{\rm eff}\sim 0.3$. Such properties are well consistent with that predicted for the $\geq 2$g merger events in the AGN disk \cite{2019PhRvL.123r1101Y}. Our results suggest that besides the dominant population consisting of the field binaries, about $6\%$ of the BBH merger events were likely from the AGN accretion disks.
\end{itemize}

In reality, the PPISN mass cutoff and the details of hierarchical black hole mergers are likely also dependent on for instance the nuclear reaction rate \cite{2019ApJ...887...53F}, stellar wind \cite{2021MNRAS.504..146V}, and accretion and detailed dynamics in the AGN disk \cite{2020ApJ...898...25T}. With a significantly extended sample in the near future, the statistical studies are expected to reliably probe these possibilities.

\begin{acknowledgments}
{\it Acknowledgments.---} This work was supported in part by NSFC under grants of No. 11921003, No. 11773078, and No. 11933010. YQ acknowledges the support from the Doctoral research start-up funding of Anhui Normal University and the funding from Key Laboratory for Relativistic Astrophysics in Guangxi University.
\end{acknowledgments}

\bibliographystyle{apsrev4-1}
\bibliography{bibliography}

\begin{thebibliography}{50}%
\makeatletter
\providecommand \@ifxundefined [1]{%
 \@ifx{#1\undefined}
}%
\providecommand \@ifnum [1]{%
 \ifnum #1\expandafter \@firstoftwo
 \else \expandafter \@secondoftwo
 \fi
}%
\providecommand \@ifx [1]{%
 \ifx #1\expandafter \@firstoftwo
 \else \expandafter \@secondoftwo
 \fi
}%
\providecommand \natexlab [1]{#1}%
\providecommand \enquote  [1]{``#1''}%
\providecommand \bibnamefont  [1]{#1}%
\providecommand \bibfnamefont [1]{#1}%
\providecommand \citenamefont [1]{#1}%
\providecommand \href@noop [0]{\@secondoftwo}%
\providecommand \href [0]{\begingroup \@sanitize@url \@href}%
\providecommand \@href[1]{\@@startlink{#1}\@@href}%
\providecommand \@@href[1]{\endgroup#1\@@endlink}%
\providecommand \@sanitize@url [0]{\catcode `\\12\catcode `\$12\catcode
  `\&12\catcode `\#12\catcode `\^12\catcode `\_12\catcode `\%12\relax}%
\providecommand \@@startlink[1]{}%
\providecommand \@@endlink[0]{}%
\providecommand \url  [0]{\begingroup\@sanitize@url \@url }%
\providecommand \@url [1]{\endgroup\@href {#1}{\urlprefix }}%
\providecommand \urlprefix  [0]{URL }%
\providecommand \Eprint [0]{\href }%
\providecommand \doibase [0]{http://dx.doi.org/}%
\providecommand \selectlanguage [0]{\@gobble}%
\providecommand \bibinfo  [0]{\@secondoftwo}%
\providecommand \bibfield  [0]{\@secondoftwo}%
\providecommand \translation [1]{[#1]}%
\providecommand \BibitemOpen [0]{}%
\providecommand \bibitemStop [0]{}%
\providecommand \bibitemNoStop [0]{.\EOS\space}%
\providecommand \EOS [0]{\spacefactor3000\relax}%
\providecommand \BibitemShut  [1]{\csname bibitem#1\endcsname}%
\let\auto@bib@innerbib\@empty
\bibitem [{\citenamefont {{Abbott}}\ \emph {et~al.}(2019)\citenamefont
  {{Abbott}}, \citenamefont {{Abbott}}, \citenamefont {{Abbott}}, \citenamefont
  {{Abraham}}, \citenamefont {{Acernese}}, \citenamefont {{Ackley}},
  \citenamefont {{Adams}}, \citenamefont {{Adhikari}}, \citenamefont {{Adya}},
  \citenamefont {{Affeldt}},\ and\ \citenamefont {et~al.}}]{abbottO2}%
  \BibitemOpen
  \bibfield  {author} {\bibinfo {author} {\bibfnamefont {B.~P.}\ \bibnamefont
  {{Abbott}}}, \bibinfo {author} {\bibfnamefont {R.}~\bibnamefont {{Abbott}}},
  \bibinfo {author} {\bibfnamefont {T.~D.}\ \bibnamefont {{Abbott}}}, \bibinfo
  {author} {\bibfnamefont {S.}~\bibnamefont {{Abraham}}}, \bibinfo {author}
  {\bibfnamefont {F.}~\bibnamefont {{Acernese}}}, \bibinfo {author}
  {\bibfnamefont {K.}~\bibnamefont {{Ackley}}}, \bibinfo {author}
  {\bibfnamefont {C.}~\bibnamefont {{Adams}}}, \bibinfo {author} {\bibfnamefont
  {R.~X.}\ \bibnamefont {{Adhikari}}}, \bibinfo {author} {\bibfnamefont
  {V.~B.}\ \bibnamefont {{Adya}}}, \bibinfo {author} {\bibfnamefont
  {C.}~\bibnamefont {{Affeldt}}}, \ and\ \bibinfo {author} {\bibnamefont
  {et~al.}},\ }\href {\doibase 10.1103/PhysRevX.9.031040} {\bibfield  {journal}
  {\bibinfo  {journal} {Physical Review X}\ }\textbf {\bibinfo {volume} {9}},\
  \bibinfo {eid} {031040} (\bibinfo {year} {2019})},\ \Eprint
  {http://arxiv.org/abs/1811.12907} {arXiv:1811.12907 [astro-ph.HE]}
  \BibitemShut {NoStop}%
\bibitem [{\citenamefont {{The LIGO Scientific Collaboration}}\ and\
  \citenamefont {{the Virgo Collaboration}}(2021)}]{2021arXiv210801045T}%
  \BibitemOpen
  \bibfield  {author} {\bibinfo {author} {\bibnamefont {{The LIGO Scientific
  Collaboration}}}\ and\ \bibinfo {author} {\bibnamefont {{the Virgo
  Collaboration}}},\ }\href@noop {} {\bibfield  {journal} {\bibinfo  {journal}
  {arXiv e-prints}\ ,\ \bibinfo {eid} {arXiv:2108.01045}} (\bibinfo {year}
  {2021})},\ \Eprint {http://arxiv.org/abs/2108.01045} {arXiv:2108.01045
  [gr-qc]} \BibitemShut {NoStop}%
\bibitem [{\citenamefont {{Mapelli}}(2021)}]{mapelli2021review}%
  \BibitemOpen
  \bibfield  {author} {\bibinfo {author} {\bibfnamefont {M.}~\bibnamefont
  {{Mapelli}}},\ }\href@noop {} {\bibfield  {journal} {\bibinfo  {journal}
  {arXiv e-prints}\ ,\ \bibinfo {eid} {arXiv:2106.00699}} (\bibinfo {year}
  {2021})},\ \Eprint {http://arxiv.org/abs/2106.00699} {arXiv:2106.00699
  [astro-ph.HE]} \BibitemShut {NoStop}%
\bibitem [{\citenamefont {{Gerosa}}\ and\ \citenamefont
  {{Fishbach}}(2021)}]{gerosa2021review}%
  \BibitemOpen
  \bibfield  {author} {\bibinfo {author} {\bibfnamefont {D.}~\bibnamefont
  {{Gerosa}}}\ and\ \bibinfo {author} {\bibfnamefont {M.}~\bibnamefont
  {{Fishbach}}},\ }\href {\doibase 10.1038/s41550-021-01398-w} {\bibfield
  {journal} {\bibinfo  {journal} {Nature Astronomy}\ } (\bibinfo {year}
  {2021}),\ 10.1038/s41550-021-01398-w},\ \Eprint
  {http://arxiv.org/abs/2105.03439} {arXiv:2105.03439 [astro-ph.HE]}
  \BibitemShut {NoStop}%
\bibitem [{\citenamefont {{Belczynski}}\ \emph
  {et~al.}(2016{\natexlab{a}})\citenamefont {{Belczynski}}, \citenamefont
  {{Holz}}, \citenamefont {{Bulik}},\ and\ \citenamefont
  {{O'Shaughnessy}}}]{2016Natur.534..512B}%
  \BibitemOpen
  \bibfield  {author} {\bibinfo {author} {\bibfnamefont {K.}~\bibnamefont
  {{Belczynski}}}, \bibinfo {author} {\bibfnamefont {D.~E.}\ \bibnamefont
  {{Holz}}}, \bibinfo {author} {\bibfnamefont {T.}~\bibnamefont {{Bulik}}}, \
  and\ \bibinfo {author} {\bibfnamefont {R.}~\bibnamefont {{O'Shaughnessy}}},\
  }\href {\doibase 10.1038/nature18322} {\bibfield  {journal} {\bibinfo
  {journal} {\nat}\ }\textbf {\bibinfo {volume} {534}},\ \bibinfo {pages} {512}
  (\bibinfo {year} {2016}{\natexlab{a}})},\ \Eprint
  {http://arxiv.org/abs/1602.04531} {arXiv:1602.04531 [astro-ph.HE]}
  \BibitemShut {NoStop}%
\bibitem [{\citenamefont {{Ivanova}}\ \emph {et~al.}(2013)\citenamefont
  {{Ivanova}}, \citenamefont {{Justham}}, \citenamefont {{Chen}}, \citenamefont
  {{De Marco}}, \citenamefont {{Fryer}}, \citenamefont {{Gaburov}},
  \citenamefont {{Ge}}, \citenamefont {{Glebbeek}}, \citenamefont {{Han}},
  \citenamefont {{Li}}, \citenamefont {{Lu}}, \citenamefont {{Marsh}},
  \citenamefont {{Podsiadlowski}}, \citenamefont {{Potter}}, \citenamefont
  {{Soker}}, \citenamefont {{Taam}}, \citenamefont {{Tauris}}, \citenamefont
  {{van den Heuvel}},\ and\ \citenamefont {{Webbink}}}]{2013A&ARv..21...59I}%
  \BibitemOpen
  \bibfield  {author} {\bibinfo {author} {\bibfnamefont {N.}~\bibnamefont
  {{Ivanova}}}, \bibinfo {author} {\bibfnamefont {S.}~\bibnamefont
  {{Justham}}}, \bibinfo {author} {\bibfnamefont {X.}~\bibnamefont {{Chen}}},
  \bibinfo {author} {\bibfnamefont {O.}~\bibnamefont {{De Marco}}}, \bibinfo
  {author} {\bibfnamefont {C.~L.}\ \bibnamefont {{Fryer}}}, \bibinfo {author}
  {\bibfnamefont {E.}~\bibnamefont {{Gaburov}}}, \bibinfo {author}
  {\bibfnamefont {H.}~\bibnamefont {{Ge}}}, \bibinfo {author} {\bibfnamefont
  {E.}~\bibnamefont {{Glebbeek}}}, \bibinfo {author} {\bibfnamefont
  {Z.}~\bibnamefont {{Han}}}, \bibinfo {author} {\bibfnamefont {X.~D.}\
  \bibnamefont {{Li}}}, \bibinfo {author} {\bibfnamefont {G.}~\bibnamefont
  {{Lu}}}, \bibinfo {author} {\bibfnamefont {T.}~\bibnamefont {{Marsh}}},
  \bibinfo {author} {\bibfnamefont {P.}~\bibnamefont {{Podsiadlowski}}},
  \bibinfo {author} {\bibfnamefont {A.}~\bibnamefont {{Potter}}}, \bibinfo
  {author} {\bibfnamefont {N.}~\bibnamefont {{Soker}}}, \bibinfo {author}
  {\bibfnamefont {R.}~\bibnamefont {{Taam}}}, \bibinfo {author} {\bibfnamefont
  {T.~M.}\ \bibnamefont {{Tauris}}}, \bibinfo {author} {\bibfnamefont
  {E.~P.~J.}\ \bibnamefont {{van den Heuvel}}}, \ and\ \bibinfo {author}
  {\bibfnamefont {R.~F.}\ \bibnamefont {{Webbink}}},\ }\href {\doibase
  10.1007/s00159-013-0059-2} {\bibfield  {journal} {\bibinfo  {journal} {aapr}\
  }\textbf {\bibinfo {volume} {21}},\ \bibinfo {eid} {59} (\bibinfo {year}
  {2013})},\ \Eprint {http://arxiv.org/abs/1209.4302} {arXiv:1209.4302
  [astro-ph.HE]} \BibitemShut {NoStop}%
\bibitem [{\citenamefont {{Belczynski}}\ \emph
  {et~al.}(2016{\natexlab{b}})\citenamefont {{Belczynski}}, \citenamefont
  {{Heger}}, \citenamefont {{Gladysz}}, \citenamefont {{Ruiter}}, \citenamefont
  {{Woosley}}, \citenamefont {{Wiktorowicz}}, \citenamefont {{Chen}},
  \citenamefont {{Bulik}}, \citenamefont {{O'Shaughnessy}}, \citenamefont
  {{Holz}}, \citenamefont {{Fryer}},\ and\ \citenamefont
  {{Berti}}}]{2016A&A...594A..97B}%
  \BibitemOpen
  \bibfield  {author} {\bibinfo {author} {\bibfnamefont {K.}~\bibnamefont
  {{Belczynski}}}, \bibinfo {author} {\bibfnamefont {A.}~\bibnamefont
  {{Heger}}}, \bibinfo {author} {\bibfnamefont {W.}~\bibnamefont {{Gladysz}}},
  \bibinfo {author} {\bibfnamefont {A.~J.}\ \bibnamefont {{Ruiter}}}, \bibinfo
  {author} {\bibfnamefont {S.}~\bibnamefont {{Woosley}}}, \bibinfo {author}
  {\bibfnamefont {G.}~\bibnamefont {{Wiktorowicz}}}, \bibinfo {author}
  {\bibfnamefont {H.~Y.}\ \bibnamefont {{Chen}}}, \bibinfo {author}
  {\bibfnamefont {T.}~\bibnamefont {{Bulik}}}, \bibinfo {author} {\bibfnamefont
  {R.}~\bibnamefont {{O'Shaughnessy}}}, \bibinfo {author} {\bibfnamefont
  {D.~E.}\ \bibnamefont {{Holz}}}, \bibinfo {author} {\bibfnamefont {C.~L.}\
  \bibnamefont {{Fryer}}}, \ and\ \bibinfo {author} {\bibfnamefont
  {E.}~\bibnamefont {{Berti}}},\ }\href {\doibase 10.1051/0004-6361/201628980}
  {\bibfield  {journal} {\bibinfo  {journal} {\aap}\ }\textbf {\bibinfo
  {volume} {594}},\ \bibinfo {eid} {A97} (\bibinfo {year}
  {2016}{\natexlab{b}})},\ \Eprint {http://arxiv.org/abs/1607.03116}
  {arXiv:1607.03116 [astro-ph.HE]} \BibitemShut {NoStop}%
\bibitem [{\citenamefont {{Woosley}}(2017)}]{2017ApJ...836..244W}%
  \BibitemOpen
  \bibfield  {author} {\bibinfo {author} {\bibfnamefont {S.~E.}\ \bibnamefont
  {{Woosley}}},\ }\href {\doibase 10.3847/1538-4357/836/2/244} {\bibfield
  {journal} {\bibinfo  {journal} {\apj}\ }\textbf {\bibinfo {volume} {836}},\
  \bibinfo {eid} {244} (\bibinfo {year} {2017})},\ \Eprint
  {http://arxiv.org/abs/1608.08939} {arXiv:1608.08939 [astro-ph.HE]}
  \BibitemShut {NoStop}%
\bibitem [{\citenamefont {{Spera}}\ and\ \citenamefont
  {{Mapelli}}(2017)}]{2017MNRAS.470.4739S}%
  \BibitemOpen
  \bibfield  {author} {\bibinfo {author} {\bibfnamefont {M.}~\bibnamefont
  {{Spera}}}\ and\ \bibinfo {author} {\bibfnamefont {M.}~\bibnamefont
  {{Mapelli}}},\ }\href {\doibase 10.1093/mnras/stx1576} {\bibfield  {journal}
  {\bibinfo  {journal} {\mnras}\ }\textbf {\bibinfo {volume} {470}},\ \bibinfo
  {pages} {4739} (\bibinfo {year} {2017})},\ \Eprint
  {http://arxiv.org/abs/1706.06109} {arXiv:1706.06109 [astro-ph.SR]}
  \BibitemShut {NoStop}%
\bibitem [{\citenamefont {{Stevenson}}\ \emph {et~al.}(2019)\citenamefont
  {{Stevenson}}, \citenamefont {{Sampson}}, \citenamefont {{Powell}},
  \citenamefont {{Vigna-G{\'o}mez}}, \citenamefont {{Neijssel}}, \citenamefont
  {{Sz{\'e}csi}},\ and\ \citenamefont {{Mandel}}}]{2019ApJ...882..121S}%
  \BibitemOpen
  \bibfield  {author} {\bibinfo {author} {\bibfnamefont {S.}~\bibnamefont
  {{Stevenson}}}, \bibinfo {author} {\bibfnamefont {M.}~\bibnamefont
  {{Sampson}}}, \bibinfo {author} {\bibfnamefont {J.}~\bibnamefont {{Powell}}},
  \bibinfo {author} {\bibfnamefont {A.}~\bibnamefont {{Vigna-G{\'o}mez}}},
  \bibinfo {author} {\bibfnamefont {C.~J.}\ \bibnamefont {{Neijssel}}},
  \bibinfo {author} {\bibfnamefont {D.}~\bibnamefont {{Sz{\'e}csi}}}, \ and\
  \bibinfo {author} {\bibfnamefont {I.}~\bibnamefont {{Mandel}}},\ }\href
  {\doibase 10.3847/1538-4357/ab3981} {\bibfield  {journal} {\bibinfo
  {journal} {\apj}\ }\textbf {\bibinfo {volume} {882}},\ \bibinfo {eid} {121}
  (\bibinfo {year} {2019})},\ \Eprint {http://arxiv.org/abs/1904.02821}
  {arXiv:1904.02821 [astro-ph.HE]} \BibitemShut {NoStop}%
\bibitem [{\citenamefont {{Mapelli}}\ \emph {et~al.}(2020)\citenamefont
  {{Mapelli}}, \citenamefont {{Spera}}, \citenamefont {{Montanari}},
  \citenamefont {{Limongi}}, \citenamefont {{Chieffi}}, \citenamefont
  {{Giacobbo}}, \citenamefont {{Bressan}},\ and\ \citenamefont
  {{Bouffanais}}}]{2020ApJ...888...76M}%
  \BibitemOpen
  \bibfield  {author} {\bibinfo {author} {\bibfnamefont {M.}~\bibnamefont
  {{Mapelli}}}, \bibinfo {author} {\bibfnamefont {M.}~\bibnamefont {{Spera}}},
  \bibinfo {author} {\bibfnamefont {E.}~\bibnamefont {{Montanari}}}, \bibinfo
  {author} {\bibfnamefont {M.}~\bibnamefont {{Limongi}}}, \bibinfo {author}
  {\bibfnamefont {A.}~\bibnamefont {{Chieffi}}}, \bibinfo {author}
  {\bibfnamefont {N.}~\bibnamefont {{Giacobbo}}}, \bibinfo {author}
  {\bibfnamefont {A.}~\bibnamefont {{Bressan}}}, \ and\ \bibinfo {author}
  {\bibfnamefont {Y.}~\bibnamefont {{Bouffanais}}},\ }\href {\doibase
  10.3847/1538-4357/ab584d} {\bibfield  {journal} {\bibinfo  {journal} {\apj}\
  }\textbf {\bibinfo {volume} {888}},\ \bibinfo {eid} {76} (\bibinfo {year}
  {2020})},\ \Eprint {http://arxiv.org/abs/1909.01371} {arXiv:1909.01371
  [astro-ph.HE]} \BibitemShut {NoStop}%
\bibitem [{\citenamefont {{Qin}}\ \emph {et~al.}(2018)\citenamefont {{Qin}},
  \citenamefont {{Fragos}}, \citenamefont {{Meynet}}, \citenamefont
  {{Andrews}}, \citenamefont {{S{\o}rensen}},\ and\ \citenamefont
  {{Song}}}]{2018A&A...616A..28Q}%
  \BibitemOpen
  \bibfield  {author} {\bibinfo {author} {\bibfnamefont {Y.}~\bibnamefont
  {{Qin}}}, \bibinfo {author} {\bibfnamefont {T.}~\bibnamefont {{Fragos}}},
  \bibinfo {author} {\bibfnamefont {G.}~\bibnamefont {{Meynet}}}, \bibinfo
  {author} {\bibfnamefont {J.}~\bibnamefont {{Andrews}}}, \bibinfo {author}
  {\bibfnamefont {M.}~\bibnamefont {{S{\o}rensen}}}, \ and\ \bibinfo {author}
  {\bibfnamefont {H.~F.}\ \bibnamefont {{Song}}},\ }\href {\doibase
  10.1051/0004-6361/201832839} {\bibfield  {journal} {\bibinfo  {journal}
  {\aap}\ }\textbf {\bibinfo {volume} {616}},\ \bibinfo {eid} {A28} (\bibinfo
  {year} {2018})},\ \Eprint {http://arxiv.org/abs/1802.05738} {arXiv:1802.05738
  [astro-ph.SR]} \BibitemShut {NoStop}%
\bibitem [{\citenamefont {{Fuller}}\ \emph {et~al.}(2019)\citenamefont
  {{Fuller}}, \citenamefont {{Piro}},\ and\ \citenamefont
  {{Jermyn}}}]{2019MNRAS.485.3661F}%
  \BibitemOpen
  \bibfield  {author} {\bibinfo {author} {\bibfnamefont {J.}~\bibnamefont
  {{Fuller}}}, \bibinfo {author} {\bibfnamefont {A.~L.}\ \bibnamefont
  {{Piro}}}, \ and\ \bibinfo {author} {\bibfnamefont {A.~S.}\ \bibnamefont
  {{Jermyn}}},\ }\href {\doibase 10.1093/mnras/stz514} {\bibfield  {journal}
  {\bibinfo  {journal} {\mnras}\ }\textbf {\bibinfo {volume} {485}},\ \bibinfo
  {pages} {3661} (\bibinfo {year} {2019})},\ \Eprint
  {http://arxiv.org/abs/1902.08227} {arXiv:1902.08227 [astro-ph.SR]}
  \BibitemShut {NoStop}%
\bibitem [{\citenamefont {{Belczynski}}\ \emph {et~al.}(2020)\citenamefont
  {{Belczynski}}, \citenamefont {{Klencki}}, \citenamefont {{Fields}},
  \citenamefont {{Olejak}}, \citenamefont {{Berti}}, \citenamefont {{Meynet}},
  \citenamefont {{Fryer}}, \citenamefont {{Holz}}, \citenamefont
  {{O'Shaughnessy}}, \citenamefont {{Brown}}, \citenamefont {{Bulik}},
  \citenamefont {{Leung}}, \citenamefont {{Nomoto}}, \citenamefont {{Madau}},
  \citenamefont {{Hirschi}}, \citenamefont {{Kaiser}}, \citenamefont {{Jones}},
  \citenamefont {{Mondal}}, \citenamefont {{Chruslinska}}, \citenamefont
  {{Drozda}}, \citenamefont {{Gerosa}}, \citenamefont {{Doctor}}, \citenamefont
  {{Giersz}}, \citenamefont {{Ekstrom}}, \citenamefont {{Georgy}},
  \citenamefont {{Askar}}, \citenamefont {{Baibhav}}, \citenamefont
  {{Wysocki}}, \citenamefont {{Natan}}, \citenamefont {{Farr}}, \citenamefont
  {{Wiktorowicz}}, \citenamefont {{Coleman Miller}}, \citenamefont {{Farr}},\
  and\ \citenamefont {{Lasota}}}]{2020A&A...636A.104B}%
  \BibitemOpen
  \bibfield  {author} {\bibinfo {author} {\bibfnamefont {K.}~\bibnamefont
  {{Belczynski}}}, \bibinfo {author} {\bibfnamefont {J.}~\bibnamefont
  {{Klencki}}}, \bibinfo {author} {\bibfnamefont {C.~E.}\ \bibnamefont
  {{Fields}}}, \bibinfo {author} {\bibfnamefont {A.}~\bibnamefont {{Olejak}}},
  \bibinfo {author} {\bibfnamefont {E.}~\bibnamefont {{Berti}}}, \bibinfo
  {author} {\bibfnamefont {G.}~\bibnamefont {{Meynet}}}, \bibinfo {author}
  {\bibfnamefont {C.~L.}\ \bibnamefont {{Fryer}}}, \bibinfo {author}
  {\bibfnamefont {D.~E.}\ \bibnamefont {{Holz}}}, \bibinfo {author}
  {\bibfnamefont {R.}~\bibnamefont {{O'Shaughnessy}}}, \bibinfo {author}
  {\bibfnamefont {D.~A.}\ \bibnamefont {{Brown}}}, \bibinfo {author}
  {\bibfnamefont {T.}~\bibnamefont {{Bulik}}}, \bibinfo {author} {\bibfnamefont
  {S.~C.}\ \bibnamefont {{Leung}}}, \bibinfo {author} {\bibfnamefont
  {K.}~\bibnamefont {{Nomoto}}}, \bibinfo {author} {\bibfnamefont
  {P.}~\bibnamefont {{Madau}}}, \bibinfo {author} {\bibfnamefont
  {R.}~\bibnamefont {{Hirschi}}}, \bibinfo {author} {\bibfnamefont
  {E.}~\bibnamefont {{Kaiser}}}, \bibinfo {author} {\bibfnamefont
  {S.}~\bibnamefont {{Jones}}}, \bibinfo {author} {\bibfnamefont
  {S.}~\bibnamefont {{Mondal}}}, \bibinfo {author} {\bibfnamefont
  {M.}~\bibnamefont {{Chruslinska}}}, \bibinfo {author} {\bibfnamefont
  {P.}~\bibnamefont {{Drozda}}}, \bibinfo {author} {\bibfnamefont
  {D.}~\bibnamefont {{Gerosa}}}, \bibinfo {author} {\bibfnamefont
  {Z.}~\bibnamefont {{Doctor}}}, \bibinfo {author} {\bibfnamefont
  {M.}~\bibnamefont {{Giersz}}}, \bibinfo {author} {\bibfnamefont
  {S.}~\bibnamefont {{Ekstrom}}}, \bibinfo {author} {\bibfnamefont
  {C.}~\bibnamefont {{Georgy}}}, \bibinfo {author} {\bibfnamefont
  {A.}~\bibnamefont {{Askar}}}, \bibinfo {author} {\bibfnamefont
  {V.}~\bibnamefont {{Baibhav}}}, \bibinfo {author} {\bibfnamefont
  {D.}~\bibnamefont {{Wysocki}}}, \bibinfo {author} {\bibfnamefont
  {T.}~\bibnamefont {{Natan}}}, \bibinfo {author} {\bibfnamefont {W.~M.}\
  \bibnamefont {{Farr}}}, \bibinfo {author} {\bibfnamefont {G.}~\bibnamefont
  {{Wiktorowicz}}}, \bibinfo {author} {\bibfnamefont {M.}~\bibnamefont
  {{Coleman Miller}}}, \bibinfo {author} {\bibfnamefont {B.}~\bibnamefont
  {{Farr}}}, \ and\ \bibinfo {author} {\bibfnamefont {J.~P.}\ \bibnamefont
  {{Lasota}}},\ }\href {\doibase 10.1051/0004-6361/201936528} {\bibfield
  {journal} {\bibinfo  {journal} {\aap}\ }\textbf {\bibinfo {volume} {636}},\
  \bibinfo {eid} {A104} (\bibinfo {year} {2020})},\ \Eprint
  {http://arxiv.org/abs/1706.07053} {arXiv:1706.07053 [astro-ph.HE]}
  \BibitemShut {NoStop}%
\bibitem [{\citenamefont {{Miller}}\ and\ \citenamefont
  {{Hamilton}}(2002)}]{miller2002}%
  \BibitemOpen
  \bibfield  {author} {\bibinfo {author} {\bibfnamefont {M.~C.}\ \bibnamefont
  {{Miller}}}\ and\ \bibinfo {author} {\bibfnamefont {D.~P.}\ \bibnamefont
  {{Hamilton}}},\ }\href {\doibase 10.1046/j.1365-8711.2002.05112.x} {\bibfield
   {journal} {\bibinfo  {journal} {\mnras}\ }\textbf {\bibinfo {volume}
  {330}},\ \bibinfo {pages} {232} (\bibinfo {year} {2002})},\ \Eprint
  {http://arxiv.org/abs/astro-ph/0106188} {astro-ph/0106188} \BibitemShut
  {NoStop}%
\bibitem [{\citenamefont {{Giersz}}\ \emph {et~al.}(2015)\citenamefont
  {{Giersz}}, \citenamefont {{Leigh}}, \citenamefont {{Hypki}}, \citenamefont
  {{L{\"u}tzgendorf}},\ and\ \citenamefont {{Askar}}}]{giersz2015}%
  \BibitemOpen
  \bibfield  {author} {\bibinfo {author} {\bibfnamefont {M.}~\bibnamefont
  {{Giersz}}}, \bibinfo {author} {\bibfnamefont {N.}~\bibnamefont {{Leigh}}},
  \bibinfo {author} {\bibfnamefont {A.}~\bibnamefont {{Hypki}}}, \bibinfo
  {author} {\bibfnamefont {N.}~\bibnamefont {{L{\"u}tzgendorf}}}, \ and\
  \bibinfo {author} {\bibfnamefont {A.}~\bibnamefont {{Askar}}},\ }\href
  {\doibase 10.1093/mnras/stv2162} {\bibfield  {journal} {\bibinfo  {journal}
  {\mnras}\ }\textbf {\bibinfo {volume} {454}},\ \bibinfo {pages} {3150}
  (\bibinfo {year} {2015})},\ \Eprint {http://arxiv.org/abs/1506.05234}
  {arXiv:1506.05234} \BibitemShut {NoStop}%
\bibitem [{\citenamefont {{Fishbach}}\ \emph {et~al.}(2017)\citenamefont
  {{Fishbach}}, \citenamefont {{Holz}},\ and\ \citenamefont
  {{Farr}}}]{fishbach2017}%
  \BibitemOpen
  \bibfield  {author} {\bibinfo {author} {\bibfnamefont {M.}~\bibnamefont
  {{Fishbach}}}, \bibinfo {author} {\bibfnamefont {D.~E.}\ \bibnamefont
  {{Holz}}}, \ and\ \bibinfo {author} {\bibfnamefont {B.}~\bibnamefont
  {{Farr}}},\ }\href {\doibase 10.3847/2041-8213/aa7045} {\bibfield  {journal}
  {\bibinfo  {journal} {\apjl}\ }\textbf {\bibinfo {volume} {840}},\ \bibinfo
  {eid} {L24} (\bibinfo {year} {2017})},\ \Eprint
  {http://arxiv.org/abs/1703.06869} {arXiv:1703.06869 [astro-ph.HE]}
  \BibitemShut {NoStop}%
\bibitem [{\citenamefont {{Gerosa}}\ and\ \citenamefont
  {{Berti}}(2017)}]{gerosa2017}%
  \BibitemOpen
  \bibfield  {author} {\bibinfo {author} {\bibfnamefont {D.}~\bibnamefont
  {{Gerosa}}}\ and\ \bibinfo {author} {\bibfnamefont {E.}~\bibnamefont
  {{Berti}}},\ }\href {\doibase 10.1103/PhysRevD.95.124046} {\bibfield
  {journal} {\bibinfo  {journal} {\prd}\ }\textbf {\bibinfo {volume} {95}},\
  \bibinfo {eid} {124046} (\bibinfo {year} {2017})},\ \Eprint
  {http://arxiv.org/abs/1703.06223} {arXiv:1703.06223 [gr-qc]} \BibitemShut
  {NoStop}%
\bibitem [{\citenamefont {{Rodriguez}}\ \emph {et~al.}(2019)\citenamefont
  {{Rodriguez}}, \citenamefont {{Zevin}}, \citenamefont {{Amaro-Seoane}},
  \citenamefont {{Chatterjee}}, \citenamefont {{Kremer}}, \citenamefont
  {{Rasio}},\ and\ \citenamefont {{Ye}}}]{rodriguez2019}%
  \BibitemOpen
  \bibfield  {author} {\bibinfo {author} {\bibfnamefont {C.~L.}\ \bibnamefont
  {{Rodriguez}}}, \bibinfo {author} {\bibfnamefont {M.}~\bibnamefont
  {{Zevin}}}, \bibinfo {author} {\bibfnamefont {P.}~\bibnamefont
  {{Amaro-Seoane}}}, \bibinfo {author} {\bibfnamefont {S.}~\bibnamefont
  {{Chatterjee}}}, \bibinfo {author} {\bibfnamefont {K.}~\bibnamefont
  {{Kremer}}}, \bibinfo {author} {\bibfnamefont {F.~A.}\ \bibnamefont
  {{Rasio}}}, \ and\ \bibinfo {author} {\bibfnamefont {C.~S.}\ \bibnamefont
  {{Ye}}},\ }\href {\doibase 10.1103/PhysRevD.100.043027} {\bibfield  {journal}
  {\bibinfo  {journal} {\prd}\ }\textbf {\bibinfo {volume} {100}},\ \bibinfo
  {eid} {043027} (\bibinfo {year} {2019})},\ \Eprint
  {http://arxiv.org/abs/1906.10260} {arXiv:1906.10260 [astro-ph.HE]}
  \BibitemShut {NoStop}%
\bibitem [{\citenamefont {{Arca Sedda}}\ \emph {et~al.}(2021)\citenamefont
  {{Arca Sedda}}, \citenamefont {{Amaro Seoane}},\ and\ \citenamefont
  {{Chen}}}]{arcasedda2021b}%
  \BibitemOpen
  \bibfield  {author} {\bibinfo {author} {\bibfnamefont {M.}~\bibnamefont
  {{Arca Sedda}}}, \bibinfo {author} {\bibfnamefont {P.}~\bibnamefont {{Amaro
  Seoane}}}, \ and\ \bibinfo {author} {\bibfnamefont {X.}~\bibnamefont
  {{Chen}}},\ }\href {\doibase 10.1051/0004-6361/202037785} {\bibfield
  {journal} {\bibinfo  {journal} {\aap}\ }\textbf {\bibinfo {volume} {652}},\
  \bibinfo {eid} {A54} (\bibinfo {year} {2021})},\ \Eprint
  {http://arxiv.org/abs/2007.13746} {arXiv:2007.13746 [astro-ph.GA]}
  \BibitemShut {NoStop}%
\bibitem [{\citenamefont {{Mapelli}}\ \emph {et~al.}(2021)\citenamefont
  {{Mapelli}}, \citenamefont {{Dall'Amico}}, \citenamefont {{Bouffanais}},
  \citenamefont {{Giacobbo}}, \citenamefont {{Arca Sedda}}, \citenamefont
  {{Artale}}, \citenamefont {{Ballone}}, \citenamefont {{Di Carlo}},
  \citenamefont {{Iorio}}, \citenamefont {{Santoliquido}},\ and\ \citenamefont
  {{Torniamenti}}}]{mapelli2021}%
  \BibitemOpen
  \bibfield  {author} {\bibinfo {author} {\bibfnamefont {M.}~\bibnamefont
  {{Mapelli}}}, \bibinfo {author} {\bibfnamefont {M.}~\bibnamefont
  {{Dall'Amico}}}, \bibinfo {author} {\bibfnamefont {Y.}~\bibnamefont
  {{Bouffanais}}}, \bibinfo {author} {\bibfnamefont {N.}~\bibnamefont
  {{Giacobbo}}}, \bibinfo {author} {\bibfnamefont {M.}~\bibnamefont {{Arca
  Sedda}}}, \bibinfo {author} {\bibfnamefont {M.~C.}\ \bibnamefont {{Artale}}},
  \bibinfo {author} {\bibfnamefont {A.}~\bibnamefont {{Ballone}}}, \bibinfo
  {author} {\bibfnamefont {U.~N.}\ \bibnamefont {{Di Carlo}}}, \bibinfo
  {author} {\bibfnamefont {G.}~\bibnamefont {{Iorio}}}, \bibinfo {author}
  {\bibfnamefont {F.}~\bibnamefont {{Santoliquido}}}, \ and\ \bibinfo {author}
  {\bibfnamefont {S.}~\bibnamefont {{Torniamenti}}},\ }\href {\doibase
  10.1093/mnras/stab1334} {\bibfield  {journal} {\bibinfo  {journal} {\mnras}\
  }\textbf {\bibinfo {volume} {505}},\ \bibinfo {pages} {339} (\bibinfo {year}
  {2021})},\ \Eprint {http://arxiv.org/abs/2103.05016} {arXiv:2103.05016
  [astro-ph.HE]} \BibitemShut {NoStop}%
\bibitem [{\citenamefont {{Yang}}\ \emph {et~al.}(2019)\citenamefont {{Yang}},
  \citenamefont {{Bartos}}, \citenamefont {{Gayathri}}, \citenamefont {{Ford}},
  \citenamefont {{Haiman}}, \citenamefont {{Klimenko}}, \citenamefont
  {{Kocsis}}, \citenamefont {{M{\'a}rka}}, \citenamefont {{M{\'a}rka}},
  \citenamefont {{McKernan}},\ and\ \citenamefont
  {{O'Shaughnessy}}}]{2019PhRvL.123r1101Y}%
  \BibitemOpen
  \bibfield  {author} {\bibinfo {author} {\bibfnamefont {Y.}~\bibnamefont
  {{Yang}}}, \bibinfo {author} {\bibfnamefont {I.}~\bibnamefont {{Bartos}}},
  \bibinfo {author} {\bibfnamefont {V.}~\bibnamefont {{Gayathri}}}, \bibinfo
  {author} {\bibfnamefont {K.~E.~S.}\ \bibnamefont {{Ford}}}, \bibinfo {author}
  {\bibfnamefont {Z.}~\bibnamefont {{Haiman}}}, \bibinfo {author}
  {\bibfnamefont {S.}~\bibnamefont {{Klimenko}}}, \bibinfo {author}
  {\bibfnamefont {B.}~\bibnamefont {{Kocsis}}}, \bibinfo {author}
  {\bibfnamefont {S.}~\bibnamefont {{M{\'a}rka}}}, \bibinfo {author}
  {\bibfnamefont {Z.}~\bibnamefont {{M{\'a}rka}}}, \bibinfo {author}
  {\bibfnamefont {B.}~\bibnamefont {{McKernan}}}, \ and\ \bibinfo {author}
  {\bibfnamefont {R.}~\bibnamefont {{O'Shaughnessy}}},\ }\href {\doibase
  10.1103/PhysRevLett.123.181101} {\bibfield  {journal} {\bibinfo  {journal}
  {\prl}\ }\textbf {\bibinfo {volume} {123}},\ \bibinfo {eid} {181101}
  (\bibinfo {year} {2019})},\ \Eprint {http://arxiv.org/abs/1906.09281}
  {arXiv:1906.09281 [astro-ph.HE]} \BibitemShut {NoStop}%
\bibitem [{\citenamefont {{Tagawa}}\ \emph {et~al.}(2021)\citenamefont
  {{Tagawa}}, \citenamefont {{Haiman}}, \citenamefont {{Bartos}}, \citenamefont
  {{Kocsis}},\ and\ \citenamefont {{Omukai}}}]{2021MNRAS.507.3362T}%
  \BibitemOpen
  \bibfield  {author} {\bibinfo {author} {\bibfnamefont {H.}~\bibnamefont
  {{Tagawa}}}, \bibinfo {author} {\bibfnamefont {Z.}~\bibnamefont {{Haiman}}},
  \bibinfo {author} {\bibfnamefont {I.}~\bibnamefont {{Bartos}}}, \bibinfo
  {author} {\bibfnamefont {B.}~\bibnamefont {{Kocsis}}}, \ and\ \bibinfo
  {author} {\bibfnamefont {K.}~\bibnamefont {{Omukai}}},\ }\href {\doibase
  10.1093/mnras/stab2315} {\bibfield  {journal} {\bibinfo  {journal} {\mnras}\
  }\textbf {\bibinfo {volume} {507}},\ \bibinfo {pages} {3362} (\bibinfo {year}
  {2021})},\ \Eprint {http://arxiv.org/abs/2104.09510} {arXiv:2104.09510
  [astro-ph.HE]} \BibitemShut {NoStop}%
\bibitem [{\citenamefont {{Kimball}}\ \emph {et~al.}(2020)\citenamefont
  {{Kimball}}, \citenamefont {{Talbot}}, \citenamefont {{Berry}}, \citenamefont
  {{Carney}}, \citenamefont {{Zevin}}, \citenamefont {{Thrane}},\ and\
  \citenamefont {{Kalogera}}}]{2020ApJ...900..177K}%
  \BibitemOpen
  \bibfield  {author} {\bibinfo {author} {\bibfnamefont {C.}~\bibnamefont
  {{Kimball}}}, \bibinfo {author} {\bibfnamefont {C.}~\bibnamefont {{Talbot}}},
  \bibinfo {author} {\bibfnamefont {C.~P.~L.}\ \bibnamefont {{Berry}}},
  \bibinfo {author} {\bibfnamefont {M.}~\bibnamefont {{Carney}}}, \bibinfo
  {author} {\bibfnamefont {M.}~\bibnamefont {{Zevin}}}, \bibinfo {author}
  {\bibfnamefont {E.}~\bibnamefont {{Thrane}}}, \ and\ \bibinfo {author}
  {\bibfnamefont {V.}~\bibnamefont {{Kalogera}}},\ }\href {\doibase
  10.3847/1538-4357/aba518} {\bibfield  {journal} {\bibinfo  {journal} {\apj}\
  }\textbf {\bibinfo {volume} {900}},\ \bibinfo {eid} {177} (\bibinfo {year}
  {2020})},\ \Eprint {http://arxiv.org/abs/2005.00023} {arXiv:2005.00023
  [astro-ph.HE]} \BibitemShut {NoStop}%
\bibitem [{\citenamefont {{Abbott}}\ \emph
  {et~al.}(2021{\natexlab{a}})\citenamefont {{Abbott}}, \citenamefont
  {{Abbott}}, \citenamefont {{Abraham}}, \citenamefont {{Acernese}},
  \citenamefont {{Ackley}}, \citenamefont {{Adams}}, \citenamefont {{Adams}},
  \citenamefont {{Adhikari}} \emph {et~al.}}]{2021ApJ...913L...7A}%
  \BibitemOpen
  \bibfield  {author} {\bibinfo {author} {\bibfnamefont {R.}~\bibnamefont
  {{Abbott}}}, \bibinfo {author} {\bibfnamefont {T.~D.}\ \bibnamefont
  {{Abbott}}}, \bibinfo {author} {\bibfnamefont {S.}~\bibnamefont {{Abraham}}},
  \bibinfo {author} {\bibfnamefont {F.}~\bibnamefont {{Acernese}}}, \bibinfo
  {author} {\bibfnamefont {K.}~\bibnamefont {{Ackley}}}, \bibinfo {author}
  {\bibfnamefont {A.}~\bibnamefont {{Adams}}}, \bibinfo {author} {\bibfnamefont
  {C.}~\bibnamefont {{Adams}}}, \bibinfo {author} {\bibfnamefont {R.~X.}\
  \bibnamefont {{Adhikari}}},  \emph {et~al.},\ }\href {\doibase
  10.3847/2041-8213/abe949} {\bibfield  {journal} {\bibinfo  {journal} {\apjl}\
  }\textbf {\bibinfo {volume} {913}},\ \bibinfo {eid} {L7} (\bibinfo {year}
  {2021}{\natexlab{a}})},\ \Eprint {http://arxiv.org/abs/2010.14533}
  {arXiv:2010.14533 [astro-ph.HE]} \BibitemShut {NoStop}%
\bibitem [{\citenamefont {{Wang}}\ \emph {et~al.}(2021)\citenamefont {{Wang}},
  \citenamefont {{Tang}}, \citenamefont {{Liang}}, \citenamefont {{Han}},
  \citenamefont {{Li}}, \citenamefont {{Jin}}, \citenamefont {{Fan}},\ and\
  \citenamefont {{Wei}}}]{2021ApJ...913...42W}%
  \BibitemOpen
  \bibfield  {author} {\bibinfo {author} {\bibfnamefont {Y.-Z.}\ \bibnamefont
  {{Wang}}}, \bibinfo {author} {\bibfnamefont {S.-P.}\ \bibnamefont {{Tang}}},
  \bibinfo {author} {\bibfnamefont {Y.-F.}\ \bibnamefont {{Liang}}}, \bibinfo
  {author} {\bibfnamefont {M.-Z.}\ \bibnamefont {{Han}}}, \bibinfo {author}
  {\bibfnamefont {X.}~\bibnamefont {{Li}}}, \bibinfo {author} {\bibfnamefont
  {Z.-P.}\ \bibnamefont {{Jin}}}, \bibinfo {author} {\bibfnamefont {Y.-Z.}\
  \bibnamefont {{Fan}}}, \ and\ \bibinfo {author} {\bibfnamefont {D.-M.}\
  \bibnamefont {{Wei}}},\ }\href {\doibase 10.3847/1538-4357/abf5df} {\bibfield
   {journal} {\bibinfo  {journal} {\apj}\ }\textbf {\bibinfo {volume} {913}},\
  \bibinfo {eid} {42} (\bibinfo {year} {2021})},\ \Eprint
  {http://arxiv.org/abs/2104.02566} {arXiv:2104.02566 [astro-ph.HE]}
  \BibitemShut {NoStop}%
\bibitem [{\citenamefont {{Baxter}}\ \emph {et~al.}(2021)\citenamefont
  {{Baxter}}, \citenamefont {{Croon}}, \citenamefont {{McDermott}},\ and\
  \citenamefont {{Sakstein}}}]{2021ApJ...916L..16B}%
  \BibitemOpen
  \bibfield  {author} {\bibinfo {author} {\bibfnamefont {E.~J.}\ \bibnamefont
  {{Baxter}}}, \bibinfo {author} {\bibfnamefont {D.}~\bibnamefont {{Croon}}},
  \bibinfo {author} {\bibfnamefont {S.~D.}\ \bibnamefont {{McDermott}}}, \ and\
  \bibinfo {author} {\bibfnamefont {J.}~\bibnamefont {{Sakstein}}},\ }\href
  {\doibase 10.3847/2041-8213/ac11fc} {\bibfield  {journal} {\bibinfo
  {journal} {\apjl}\ }\textbf {\bibinfo {volume} {916}},\ \bibinfo {eid} {L16}
  (\bibinfo {year} {2021})},\ \Eprint {http://arxiv.org/abs/2104.02685}
  {arXiv:2104.02685 [astro-ph.CO]} \BibitemShut {NoStop}%
\bibitem [{\citenamefont {{Bavera}}\ \emph {et~al.}(2020)\citenamefont
  {{Bavera}}, \citenamefont {{Fragos}}, \citenamefont {{Qin}}, \citenamefont
  {{Zapartas}}, \citenamefont {{Neijssel}}, \citenamefont {{Mandel}},
  \citenamefont {{Batta}}, \citenamefont {{Gaebel}}, \citenamefont
  {{Kimball}},\ and\ \citenamefont {{Stevenson}}}]{2020A&A...635A..97B}%
  \BibitemOpen
  \bibfield  {author} {\bibinfo {author} {\bibfnamefont {S.~S.}\ \bibnamefont
  {{Bavera}}}, \bibinfo {author} {\bibfnamefont {T.}~\bibnamefont {{Fragos}}},
  \bibinfo {author} {\bibfnamefont {Y.}~\bibnamefont {{Qin}}}, \bibinfo
  {author} {\bibfnamefont {E.}~\bibnamefont {{Zapartas}}}, \bibinfo {author}
  {\bibfnamefont {C.~J.}\ \bibnamefont {{Neijssel}}}, \bibinfo {author}
  {\bibfnamefont {I.}~\bibnamefont {{Mandel}}}, \bibinfo {author}
  {\bibfnamefont {A.}~\bibnamefont {{Batta}}}, \bibinfo {author} {\bibfnamefont
  {S.~M.}\ \bibnamefont {{Gaebel}}}, \bibinfo {author} {\bibfnamefont
  {C.}~\bibnamefont {{Kimball}}}, \ and\ \bibinfo {author} {\bibfnamefont
  {S.}~\bibnamefont {{Stevenson}}},\ }\href {\doibase
  10.1051/0004-6361/201936204} {\bibfield  {journal} {\bibinfo  {journal}
  {\aap}\ }\textbf {\bibinfo {volume} {635}},\ \bibinfo {eid} {A97} (\bibinfo
  {year} {2020})},\ \Eprint {http://arxiv.org/abs/1906.12257} {arXiv:1906.12257
  [astro-ph.HE]} \BibitemShut {NoStop}%
\bibitem [{\citenamefont {{Li}}\ \emph {et~al.}(2021)\citenamefont {{Li}},
  \citenamefont {{Wang}}, \citenamefont {{Han}}, \citenamefont {{Tang}},
  \citenamefont {{Yuan}}, \citenamefont {{Fan}},\ and\ \citenamefont
  {{Wei}}}]{2021ApJ...917...33L}%
  \BibitemOpen
  \bibfield  {author} {\bibinfo {author} {\bibfnamefont {Y.-J.}\ \bibnamefont
  {{Li}}}, \bibinfo {author} {\bibfnamefont {Y.-Z.}\ \bibnamefont {{Wang}}},
  \bibinfo {author} {\bibfnamefont {M.-Z.}\ \bibnamefont {{Han}}}, \bibinfo
  {author} {\bibfnamefont {S.-P.}\ \bibnamefont {{Tang}}}, \bibinfo {author}
  {\bibfnamefont {Q.}~\bibnamefont {{Yuan}}}, \bibinfo {author} {\bibfnamefont
  {Y.-Z.}\ \bibnamefont {{Fan}}}, \ and\ \bibinfo {author} {\bibfnamefont
  {D.-M.}\ \bibnamefont {{Wei}}},\ }\href {\doibase 10.3847/1538-4357/ac0971}
  {\bibfield  {journal} {\bibinfo  {journal} {\apj}\ }\textbf {\bibinfo
  {volume} {917}},\ \bibinfo {eid} {33} (\bibinfo {year} {2021})},\ \Eprint
  {http://arxiv.org/abs/2104.02969} {arXiv:2104.02969 [astro-ph.HE]}
  \BibitemShut {NoStop}%
\bibitem [{\citenamefont {{Rinaldi}}\ and\ \citenamefont {{Del
  Pozzo}}(2021)}]{2021arXiv210905960R}%
  \BibitemOpen
  \bibfield  {author} {\bibinfo {author} {\bibfnamefont {S.}~\bibnamefont
  {{Rinaldi}}}\ and\ \bibinfo {author} {\bibfnamefont {W.}~\bibnamefont {{Del
  Pozzo}}},\ }\href@noop {} {\bibfield  {journal} {\bibinfo  {journal} {arXiv
  e-prints}\ ,\ \bibinfo {eid} {arXiv:2109.05960}} (\bibinfo {year} {2021})},\
  \Eprint {http://arxiv.org/abs/2109.05960} {arXiv:2109.05960 [astro-ph.IM]}
  \BibitemShut {NoStop}%
\bibitem [{\citenamefont {{Tiwari}}(2021)}]{2021CQGra..38o5007T}%
  \BibitemOpen
  \bibfield  {author} {\bibinfo {author} {\bibfnamefont {V.}~\bibnamefont
  {{Tiwari}}},\ }\href {\doibase 10.1088/1361-6382/ac0b54} {\bibfield
  {journal} {\bibinfo  {journal} {Classical and Quantum Gravity}\ }\textbf
  {\bibinfo {volume} {38}},\ \bibinfo {eid} {155007} (\bibinfo {year}
  {2021})},\ \Eprint {http://arxiv.org/abs/2006.15047} {arXiv:2006.15047
  [astro-ph.HE]} \BibitemShut {NoStop}%
\bibitem [{\citenamefont {{Callister}}\ \emph {et~al.}(2021)\citenamefont
  {{Callister}}, \citenamefont {{Haster}}, \citenamefont {{Ng}}, \citenamefont
  {{Vitale}},\ and\ \citenamefont {{Farr}}}]{2021arXiv210600521C}%
  \BibitemOpen
  \bibfield  {author} {\bibinfo {author} {\bibfnamefont {T.~A.}\ \bibnamefont
  {{Callister}}}, \bibinfo {author} {\bibfnamefont {C.-J.}\ \bibnamefont
  {{Haster}}}, \bibinfo {author} {\bibfnamefont {K.~K.~Y.}\ \bibnamefont
  {{Ng}}}, \bibinfo {author} {\bibfnamefont {S.}~\bibnamefont {{Vitale}}}, \
  and\ \bibinfo {author} {\bibfnamefont {W.~M.}\ \bibnamefont {{Farr}}},\
  }\href@noop {} {\bibfield  {journal} {\bibinfo  {journal} {arXiv e-prints}\
  ,\ \bibinfo {eid} {arXiv:2106.00521}} (\bibinfo {year} {2021})},\ \Eprint
  {http://arxiv.org/abs/2106.00521} {arXiv:2106.00521 [astro-ph.HE]}
  \BibitemShut {NoStop}%
\bibitem [{\citenamefont {{Safarzadeh}}\ and\ \citenamefont
  {{Wysocki}}(2021)}]{2021ApJ...907L..24S}%
  \BibitemOpen
  \bibfield  {author} {\bibinfo {author} {\bibfnamefont {M.}~\bibnamefont
  {{Safarzadeh}}}\ and\ \bibinfo {author} {\bibfnamefont {D.}~\bibnamefont
  {{Wysocki}}},\ }\href {\doibase 10.3847/2041-8213/abd8c7} {\bibfield
  {journal} {\bibinfo  {journal} {\apjl}\ }\textbf {\bibinfo {volume} {907}},\
  \bibinfo {eid} {L24} (\bibinfo {year} {2021})},\ \Eprint
  {http://arxiv.org/abs/2011.09959} {arXiv:2011.09959 [astro-ph.HE]}
  \BibitemShut {NoStop}%
\bibitem [{\citenamefont {{Tang}}\ \emph {et~al.}(2021)\citenamefont {{Tang}},
  \citenamefont {{Li}}, \citenamefont {{Wang}}, \citenamefont {{Fan}},\ and\
  \citenamefont {{Wei}}}]{2021arXiv210708811T}%
  \BibitemOpen
  \bibfield  {author} {\bibinfo {author} {\bibfnamefont {S.-P.}\ \bibnamefont
  {{Tang}}}, \bibinfo {author} {\bibfnamefont {Y.-J.}\ \bibnamefont {{Li}}},
  \bibinfo {author} {\bibfnamefont {Y.-Z.}\ \bibnamefont {{Wang}}}, \bibinfo
  {author} {\bibfnamefont {Y.-Z.}\ \bibnamefont {{Fan}}}, \ and\ \bibinfo
  {author} {\bibfnamefont {D.-M.}\ \bibnamefont {{Wei}}},\ }\href@noop {}
  {\bibfield  {journal} {\bibinfo  {journal} {arXiv e-prints}\ ,\ \bibinfo
  {eid} {arXiv:2107.08811}} (\bibinfo {year} {2021})},\ \Eprint
  {http://arxiv.org/abs/2107.08811} {arXiv:2107.08811 [astro-ph.HE]}
  \BibitemShut {NoStop}%
\bibitem [{\citenamefont {{Bouffanais}}\ \emph {et~al.}(2021)\citenamefont
  {{Bouffanais}}, \citenamefont {{Mapelli}}, \citenamefont {{Santoliquido}},
  \citenamefont {{Giacobbo}}, \citenamefont {{Di Carlo}}, \citenamefont
  {{Rastello}}, \citenamefont {{Artale}},\ and\ \citenamefont
  {{Iorio}}}]{bouffanais2021}%
  \BibitemOpen
  \bibfield  {author} {\bibinfo {author} {\bibfnamefont {Y.}~\bibnamefont
  {{Bouffanais}}}, \bibinfo {author} {\bibfnamefont {M.}~\bibnamefont
  {{Mapelli}}}, \bibinfo {author} {\bibfnamefont {F.}~\bibnamefont
  {{Santoliquido}}}, \bibinfo {author} {\bibfnamefont {N.}~\bibnamefont
  {{Giacobbo}}}, \bibinfo {author} {\bibfnamefont {U.~N.}\ \bibnamefont {{Di
  Carlo}}}, \bibinfo {author} {\bibfnamefont {S.}~\bibnamefont {{Rastello}}},
  \bibinfo {author} {\bibfnamefont {M.~C.}\ \bibnamefont {{Artale}}}, \ and\
  \bibinfo {author} {\bibfnamefont {G.}~\bibnamefont {{Iorio}}},\ }\href
  {\doibase 10.1093/mnras/stab2438} {\bibfield  {journal} {\bibinfo  {journal}
  {\mnras}\ } (\bibinfo {year} {2021}),\ 10.1093/mnras/stab2438},\ \Eprint
  {http://arxiv.org/abs/2102.12495} {arXiv:2102.12495 [astro-ph.HE]}
  \BibitemShut {NoStop}%
\bibitem [{\citenamefont {{Zevin}}\ \emph {et~al.}(2021)\citenamefont
  {{Zevin}}, \citenamefont {{Bavera}}, \citenamefont {{Berry}}, \citenamefont
  {{Kalogera}}, \citenamefont {{Fragos}}, \citenamefont {{Marchant}},
  \citenamefont {{Rodriguez}}, \citenamefont {{Antonini}}, \citenamefont
  {{Holz}},\ and\ \citenamefont {{Pankow}}}]{zevin2021}%
  \BibitemOpen
  \bibfield  {author} {\bibinfo {author} {\bibfnamefont {M.}~\bibnamefont
  {{Zevin}}}, \bibinfo {author} {\bibfnamefont {S.~S.}\ \bibnamefont
  {{Bavera}}}, \bibinfo {author} {\bibfnamefont {C.~P.~L.}\ \bibnamefont
  {{Berry}}}, \bibinfo {author} {\bibfnamefont {V.}~\bibnamefont {{Kalogera}}},
  \bibinfo {author} {\bibfnamefont {T.}~\bibnamefont {{Fragos}}}, \bibinfo
  {author} {\bibfnamefont {P.}~\bibnamefont {{Marchant}}}, \bibinfo {author}
  {\bibfnamefont {C.~L.}\ \bibnamefont {{Rodriguez}}}, \bibinfo {author}
  {\bibfnamefont {F.}~\bibnamefont {{Antonini}}}, \bibinfo {author}
  {\bibfnamefont {D.~E.}\ \bibnamefont {{Holz}}}, \ and\ \bibinfo {author}
  {\bibfnamefont {C.}~\bibnamefont {{Pankow}}},\ }\href {\doibase
  10.3847/1538-4357/abe40e} {\bibfield  {journal} {\bibinfo  {journal} {\apj}\
  }\textbf {\bibinfo {volume} {910}},\ \bibinfo {eid} {152} (\bibinfo {year}
  {2021})},\ \Eprint {http://arxiv.org/abs/2011.10057} {arXiv:2011.10057
  [astro-ph.HE]} \BibitemShut {NoStop}%
\bibitem [{\citenamefont {{Wong}}\ \emph {et~al.}(2021)\citenamefont {{Wong}},
  \citenamefont {{Breivik}}, \citenamefont {{Kremer}},\ and\ \citenamefont
  {{Callister}}}]{wong2021}%
  \BibitemOpen
  \bibfield  {author} {\bibinfo {author} {\bibfnamefont {K.~W.~K.}\
  \bibnamefont {{Wong}}}, \bibinfo {author} {\bibfnamefont {K.}~\bibnamefont
  {{Breivik}}}, \bibinfo {author} {\bibfnamefont {K.}~\bibnamefont {{Kremer}}},
  \ and\ \bibinfo {author} {\bibfnamefont {T.}~\bibnamefont {{Callister}}},\
  }\href {\doibase 10.1103/PhysRevD.103.083021} {\bibfield  {journal} {\bibinfo
   {journal} {\prd}\ }\textbf {\bibinfo {volume} {103}},\ \bibinfo {eid}
  {083021} (\bibinfo {year} {2021})},\ \Eprint
  {http://arxiv.org/abs/2011.03564} {arXiv:2011.03564 [astro-ph.HE]}
  \BibitemShut {NoStop}%
\bibitem [{\citenamefont {{Roulet}}\ \emph {et~al.}(2021)\citenamefont
  {{Roulet}}, \citenamefont {{Chia}}, \citenamefont {{Olsen}}, \citenamefont
  {{Dai}}, \citenamefont {{Venumadhav}}, \citenamefont {{Zackay}},\ and\
  \citenamefont {{Zaldarriaga}}}]{roulet2021}%
  \BibitemOpen
  \bibfield  {author} {\bibinfo {author} {\bibfnamefont {J.}~\bibnamefont
  {{Roulet}}}, \bibinfo {author} {\bibfnamefont {H.~S.}\ \bibnamefont
  {{Chia}}}, \bibinfo {author} {\bibfnamefont {S.}~\bibnamefont {{Olsen}}},
  \bibinfo {author} {\bibfnamefont {L.}~\bibnamefont {{Dai}}}, \bibinfo
  {author} {\bibfnamefont {T.}~\bibnamefont {{Venumadhav}}}, \bibinfo {author}
  {\bibfnamefont {B.}~\bibnamefont {{Zackay}}}, \ and\ \bibinfo {author}
  {\bibfnamefont {M.}~\bibnamefont {{Zaldarriaga}}},\ }\href@noop {} {\bibfield
   {journal} {\bibinfo  {journal} {arXiv e-prints}\ ,\ \bibinfo {eid}
  {arXiv:2105.10580}} (\bibinfo {year} {2021})},\ \Eprint
  {http://arxiv.org/abs/2105.10580} {arXiv:2105.10580 [astro-ph.HE]}
  \BibitemShut {NoStop}%
\bibitem [{\citenamefont {{Spruit}}(2002)}]{2002A&A...381..923S}%
  \BibitemOpen
  \bibfield  {author} {\bibinfo {author} {\bibfnamefont {H.~C.}\ \bibnamefont
  {{Spruit}}},\ }\href {\doibase 10.1051/0004-6361:20011465} {\bibfield
  {journal} {\bibinfo  {journal} {\aap}\ }\textbf {\bibinfo {volume} {381}},\
  \bibinfo {pages} {923} (\bibinfo {year} {2002})},\ \Eprint
  {http://arxiv.org/abs/astro-ph/0108207} {arXiv:astro-ph/0108207 [astro-ph]}
  \BibitemShut {NoStop}%
\bibitem [{\citenamefont {{Abbott}}\ \emph
  {et~al.}(2021{\natexlab{b}})\citenamefont {{Abbott}}, \citenamefont
  {{Abbott}}, \citenamefont {{Abraham}}, \citenamefont {{Acernese}},
  \citenamefont {{Ackley}}, \citenamefont {{Adams}}, \citenamefont {{Adams}},
  \citenamefont {{Adhikari}}, \citenamefont {{Adya}}, \citenamefont
  {{Affeldt}}, \citenamefont {{Agathos}}, \citenamefont {{Agatsuma}},
  \citenamefont {{Aggarwal}}, \citenamefont {{Aguiar}}, \citenamefont
  {{Aiello}}, \citenamefont {{Ain}}, \citenamefont {{Ajith}},\ and\
  \citenamefont {et~al.}}]{abbottO3popandrate}%
  \BibitemOpen
  \bibfield  {author} {\bibinfo {author} {\bibfnamefont {R.}~\bibnamefont
  {{Abbott}}}, \bibinfo {author} {\bibfnamefont {T.~D.}\ \bibnamefont
  {{Abbott}}}, \bibinfo {author} {\bibfnamefont {S.}~\bibnamefont {{Abraham}}},
  \bibinfo {author} {\bibfnamefont {F.}~\bibnamefont {{Acernese}}}, \bibinfo
  {author} {\bibfnamefont {K.}~\bibnamefont {{Ackley}}}, \bibinfo {author}
  {\bibfnamefont {A.}~\bibnamefont {{Adams}}}, \bibinfo {author} {\bibfnamefont
  {C.}~\bibnamefont {{Adams}}}, \bibinfo {author} {\bibfnamefont {R.~X.}\
  \bibnamefont {{Adhikari}}}, \bibinfo {author} {\bibfnamefont {V.~B.}\
  \bibnamefont {{Adya}}}, \bibinfo {author} {\bibfnamefont {C.}~\bibnamefont
  {{Affeldt}}}, \bibinfo {author} {\bibfnamefont {M.}~\bibnamefont
  {{Agathos}}}, \bibinfo {author} {\bibfnamefont {K.}~\bibnamefont
  {{Agatsuma}}}, \bibinfo {author} {\bibfnamefont {N.}~\bibnamefont
  {{Aggarwal}}}, \bibinfo {author} {\bibfnamefont {O.~D.}\ \bibnamefont
  {{Aguiar}}}, \bibinfo {author} {\bibfnamefont {L.}~\bibnamefont {{Aiello}}},
  \bibinfo {author} {\bibfnamefont {A.}~\bibnamefont {{Ain}}}, \bibinfo
  {author} {\bibfnamefont {P.}~\bibnamefont {{Ajith}}}, \ and\ \bibinfo
  {author} {\bibnamefont {et~al.}},\ }\href {\doibase 10.3847/2041-8213/abe949}
  {\bibfield  {journal} {\bibinfo  {journal} {\apjl}\ }\textbf {\bibinfo
  {volume} {913}},\ \bibinfo {eid} {L7} (\bibinfo {year}
  {2021}{\natexlab{b}})},\ \Eprint {http://arxiv.org/abs/2010.14533}
  {arXiv:2010.14533 [astro-ph.HE]} \BibitemShut {NoStop}%
\bibitem [{\citenamefont {{Talbot}}\ and\ \citenamefont
  {{Thrane}}(2018)}]{2018ApJ...856..173T}%
  \BibitemOpen
  \bibfield  {author} {\bibinfo {author} {\bibfnamefont {C.}~\bibnamefont
  {{Talbot}}}\ and\ \bibinfo {author} {\bibfnamefont {E.}~\bibnamefont
  {{Thrane}}},\ }\href {\doibase 10.3847/1538-4357/aab34c} {\bibfield
  {journal} {\bibinfo  {journal} {\apj}\ }\textbf {\bibinfo {volume} {856}},\
  \bibinfo {eid} {173} (\bibinfo {year} {2018})},\ \Eprint
  {http://arxiv.org/abs/1801.02699} {arXiv:1801.02699 [astro-ph.HE]}
  \BibitemShut {NoStop}%
\bibitem [{\citenamefont {{Abbott}}\ \emph
  {et~al.}(2021{\natexlab{c}})\citenamefont {{Abbott}}, \citenamefont
  {{Abbott}}, \citenamefont {{Abraham}}, \citenamefont {{Acernese}},
  \citenamefont {{Ackley}}, \citenamefont {{Adams}}, \citenamefont {{Adams}},
  \citenamefont {{Adhikari}}, \citenamefont {{Adya}}, \citenamefont
  {{Affeldt}}, \citenamefont {{Agathos}}, \citenamefont {{Agatsuma}},
  \citenamefont {{Aggarwal}}, \citenamefont {{Aguiar}}, \citenamefont
  {{Aiello}}, \citenamefont {{Ain}}, \citenamefont {{Ajith}},\ and\
  \citenamefont {et~al.}}]{abbottO3a}%
  \BibitemOpen
  \bibfield  {author} {\bibinfo {author} {\bibfnamefont {R.}~\bibnamefont
  {{Abbott}}}, \bibinfo {author} {\bibfnamefont {T.~D.}\ \bibnamefont
  {{Abbott}}}, \bibinfo {author} {\bibfnamefont {S.}~\bibnamefont {{Abraham}}},
  \bibinfo {author} {\bibfnamefont {F.}~\bibnamefont {{Acernese}}}, \bibinfo
  {author} {\bibfnamefont {K.}~\bibnamefont {{Ackley}}}, \bibinfo {author}
  {\bibfnamefont {A.}~\bibnamefont {{Adams}}}, \bibinfo {author} {\bibfnamefont
  {C.}~\bibnamefont {{Adams}}}, \bibinfo {author} {\bibfnamefont {R.~X.}\
  \bibnamefont {{Adhikari}}}, \bibinfo {author} {\bibfnamefont {V.~B.}\
  \bibnamefont {{Adya}}}, \bibinfo {author} {\bibfnamefont {C.}~\bibnamefont
  {{Affeldt}}}, \bibinfo {author} {\bibfnamefont {M.}~\bibnamefont
  {{Agathos}}}, \bibinfo {author} {\bibfnamefont {K.}~\bibnamefont
  {{Agatsuma}}}, \bibinfo {author} {\bibfnamefont {N.}~\bibnamefont
  {{Aggarwal}}}, \bibinfo {author} {\bibfnamefont {O.~D.}\ \bibnamefont
  {{Aguiar}}}, \bibinfo {author} {\bibfnamefont {L.}~\bibnamefont {{Aiello}}},
  \bibinfo {author} {\bibfnamefont {A.}~\bibnamefont {{Ain}}}, \bibinfo
  {author} {\bibfnamefont {P.}~\bibnamefont {{Ajith}}}, \ and\ \bibinfo
  {author} {\bibnamefont {et~al.}},\ }\href {\doibase
  10.1103/PhysRevX.11.021053} {\bibfield  {journal} {\bibinfo  {journal}
  {Physical Review X}\ }\textbf {\bibinfo {volume} {11}},\ \bibinfo {eid}
  {021053} (\bibinfo {year} {2021}{\natexlab{c}})},\ \Eprint
  {http://arxiv.org/abs/2010.14527} {arXiv:2010.14527 [gr-qc]} \BibitemShut
  {NoStop}%
\bibitem [{\citenamefont {{Thrane}}\ and\ \citenamefont
  {{Talbot}}(2019)}]{2019PASA...36...10T}%
  \BibitemOpen
  \bibfield  {author} {\bibinfo {author} {\bibfnamefont {E.}~\bibnamefont
  {{Thrane}}}\ and\ \bibinfo {author} {\bibfnamefont {C.}~\bibnamefont
  {{Talbot}}},\ }\href {\doibase 10.1017/pasa.2019.2} {\bibfield  {journal}
  {\bibinfo  {journal} {PASA}\ }\textbf {\bibinfo {volume} {36}},\ \bibinfo
  {eid} {e010} (\bibinfo {year} {2019})},\ \Eprint
  {http://arxiv.org/abs/1809.02293} {arXiv:1809.02293 [astro-ph.IM]}
  \BibitemShut {NoStop}%
\bibitem [{\citenamefont {{Galaudage}}\ \emph {et~al.}(2021)\citenamefont
  {{Galaudage}}, \citenamefont {{Talbot}}, \citenamefont {{Nagar}},
  \citenamefont {{Jain}}, \citenamefont {{Thrane}},\ and\ \citenamefont
  {{Mandel}}}]{2021arXiv210902424G}%
  \BibitemOpen
  \bibfield  {author} {\bibinfo {author} {\bibfnamefont {S.}~\bibnamefont
  {{Galaudage}}}, \bibinfo {author} {\bibfnamefont {C.}~\bibnamefont
  {{Talbot}}}, \bibinfo {author} {\bibfnamefont {T.}~\bibnamefont {{Nagar}}},
  \bibinfo {author} {\bibfnamefont {D.}~\bibnamefont {{Jain}}}, \bibinfo
  {author} {\bibfnamefont {E.}~\bibnamefont {{Thrane}}}, \ and\ \bibinfo
  {author} {\bibfnamefont {I.}~\bibnamefont {{Mandel}}},\ }\href@noop {}
  {\bibfield  {journal} {\bibinfo  {journal} {arXiv e-prints}\ ,\ \bibinfo
  {eid} {arXiv:2109.02424}} (\bibinfo {year} {2021})},\ \Eprint
  {http://arxiv.org/abs/2109.02424} {arXiv:2109.02424 [gr-qc]} \BibitemShut
  {NoStop}%
\bibitem [{\citenamefont {{Fishbach}}\ \emph {et~al.}(2020)\citenamefont
  {{Fishbach}}, \citenamefont {{Farr}},\ and\ \citenamefont
  {{Holz}}}]{2020ApJ...891L..31F}%
  \BibitemOpen
  \bibfield  {author} {\bibinfo {author} {\bibfnamefont {M.}~\bibnamefont
  {{Fishbach}}}, \bibinfo {author} {\bibfnamefont {W.~M.}\ \bibnamefont
  {{Farr}}}, \ and\ \bibinfo {author} {\bibfnamefont {D.~E.}\ \bibnamefont
  {{Holz}}},\ }\href {\doibase 10.3847/2041-8213/ab77c9} {\bibfield  {journal}
  {\bibinfo  {journal} {\apjl}\ }\textbf {\bibinfo {volume} {891}},\ \bibinfo
  {eid} {L31} (\bibinfo {year} {2020})},\ \Eprint
  {http://arxiv.org/abs/1911.05882} {arXiv:1911.05882 [astro-ph.HE]}
  \BibitemShut {NoStop}%
\bibitem [{\citenamefont {{Miller}}\ \emph {et~al.}(2020)\citenamefont
  {{Miller}}, \citenamefont {{Callister}},\ and\ \citenamefont
  {{Farr}}}]{2020ApJ...895..128M}%
  \BibitemOpen
  \bibfield  {author} {\bibinfo {author} {\bibfnamefont {S.}~\bibnamefont
  {{Miller}}}, \bibinfo {author} {\bibfnamefont {T.~A.}\ \bibnamefont
  {{Callister}}}, \ and\ \bibinfo {author} {\bibfnamefont {W.~M.}\ \bibnamefont
  {{Farr}}},\ }\href {\doibase 10.3847/1538-4357/ab80c0} {\bibfield  {journal}
  {\bibinfo  {journal} {\apj}\ }\textbf {\bibinfo {volume} {895}},\ \bibinfo
  {eid} {128} (\bibinfo {year} {2020})},\ \Eprint
  {http://arxiv.org/abs/2001.06051} {arXiv:2001.06051 [astro-ph.HE]}
  \BibitemShut {NoStop}%
\bibitem [{\citenamefont {{Tagawa}}\ \emph {et~al.}(2020)\citenamefont
  {{Tagawa}}, \citenamefont {{Haiman}},\ and\ \citenamefont
  {{Kocsis}}}]{2020ApJ...898...25T}%
  \BibitemOpen
  \bibfield  {author} {\bibinfo {author} {\bibfnamefont {H.}~\bibnamefont
  {{Tagawa}}}, \bibinfo {author} {\bibfnamefont {Z.}~\bibnamefont {{Haiman}}},
  \ and\ \bibinfo {author} {\bibfnamefont {B.}~\bibnamefont {{Kocsis}}},\
  }\href {\doibase 10.3847/1538-4357/ab9b8c} {\bibfield  {journal} {\bibinfo
  {journal} {\apj}\ }\textbf {\bibinfo {volume} {898}},\ \bibinfo {eid} {25}
  (\bibinfo {year} {2020})},\ \Eprint {http://arxiv.org/abs/1912.08218}
  {arXiv:1912.08218 [astro-ph.GA]} \BibitemShut {NoStop}%
\bibitem [{\citenamefont {{Yi}}\ and\ \citenamefont
  {{Cheng}}(2019)}]{2019ApJ...884L..12Y}%
  \BibitemOpen
  \bibfield  {author} {\bibinfo {author} {\bibfnamefont {S.-X.}\ \bibnamefont
  {{Yi}}}\ and\ \bibinfo {author} {\bibfnamefont {K.~S.}\ \bibnamefont
  {{Cheng}}},\ }\href {\doibase 10.3847/2041-8213/ab459a} {\bibfield  {journal}
  {\bibinfo  {journal} {\apjl}\ }\textbf {\bibinfo {volume} {884}},\ \bibinfo
  {eid} {L12} (\bibinfo {year} {2019})},\ \Eprint
  {http://arxiv.org/abs/1909.08384} {arXiv:1909.08384 [astro-ph.HE]}
  \BibitemShut {NoStop}%
\bibitem [{\citenamefont {{Farmer}}\ \emph {et~al.}(2019)\citenamefont
  {{Farmer}}, \citenamefont {{Renzo}}, \citenamefont {{de Mink}}, \citenamefont
  {{Marchant}},\ and\ \citenamefont {{Justham}}}]{2019ApJ...887...53F}%
  \BibitemOpen
  \bibfield  {author} {\bibinfo {author} {\bibfnamefont {R.}~\bibnamefont
  {{Farmer}}}, \bibinfo {author} {\bibfnamefont {M.}~\bibnamefont {{Renzo}}},
  \bibinfo {author} {\bibfnamefont {S.~E.}\ \bibnamefont {{de Mink}}}, \bibinfo
  {author} {\bibfnamefont {P.}~\bibnamefont {{Marchant}}}, \ and\ \bibinfo
  {author} {\bibfnamefont {S.}~\bibnamefont {{Justham}}},\ }\href {\doibase
  10.3847/1538-4357/ab518b} {\bibfield  {journal} {\bibinfo  {journal} {\apj}\
  }\textbf {\bibinfo {volume} {887}},\ \bibinfo {eid} {53} (\bibinfo {year}
  {2019})},\ \Eprint {http://arxiv.org/abs/1910.12874} {arXiv:1910.12874
  [astro-ph.SR]} \BibitemShut {NoStop}%
\bibitem [{\citenamefont {{Vink}}\ \emph {et~al.}(2021)\citenamefont {{Vink}},
  \citenamefont {{Higgins}}, \citenamefont {{Sander}},\ and\ \citenamefont
  {{Sabhahit}}}]{2021MNRAS.504..146V}%
  \BibitemOpen
  \bibfield  {author} {\bibinfo {author} {\bibfnamefont {J.~S.}\ \bibnamefont
  {{Vink}}}, \bibinfo {author} {\bibfnamefont {E.~R.}\ \bibnamefont
  {{Higgins}}}, \bibinfo {author} {\bibfnamefont {A.~A.~C.}\ \bibnamefont
  {{Sander}}}, \ and\ \bibinfo {author} {\bibfnamefont {G.~N.}\ \bibnamefont
  {{Sabhahit}}},\ }\href {\doibase 10.1093/mnras/stab842} {\bibfield  {journal}
  {\bibinfo  {journal} {\mnras}\ }\textbf {\bibinfo {volume} {504}},\ \bibinfo
  {pages} {146} (\bibinfo {year} {2021})},\ \Eprint
  {http://arxiv.org/abs/2010.11730} {arXiv:2010.11730 [astro-ph.HE]}
  \BibitemShut {NoStop}%
\end{thebibliography}%

\end{document}